\documentclass[11pt]{article}

\usepackage{acl}

\usepackage{times}
\usepackage{latexsym}

\usepackage[utf8]{inputenc}
\usepackage{amsmath}

\usepackage{tabularx}
\usepackage{pgfplots}
\pgfplotsset{compat=1.18}
\usepgfplotslibrary{polar, statistics}

\usepackage{booktabs}
\usepackage{array}
\usepackage{longtable}
\usepackage{multirow}
\usepackage{makecell}

\usepackage[table]{xcolor}
\usepackage{tcolorbox}

\usepackage{listings}
\usepackage{enumitem}
\usepackage{graphicx}

\definecolor{freshblue}{RGB}{66, 165, 245}
\definecolor{softblue}{RGB}{227, 242, 253}

\definecolor{freshgreen}{RGB}{102, 187, 106}
\definecolor{softgreen}{RGB}{232, 245, 233}

\definecolor{mygray}{gray}{0.95}
\definecolor{myblue}{RGB}{235, 245, 255}
\definecolor{wincolor}{RGB}{178, 34, 34}

\newcommand{\best}[1]{\textbf{#1}}
\tcbuselibrary{listings,breakable}

\usepackage{xfp}
\definecolor{MyBlue}{HTML}{E6F0FF}
\definecolor{MyGreen}{HTML}{E6FFFA}
\definecolor{MyRed}{HTML}{FFF0F0}

\usepackage{microtype}
\usepackage{inconsolata}

\title{SecGoal: A Benchmark for Extracting Formalizable Security Goals from Protocol Documents}

\author{
  Dawei Huang, Hui Li\thanks{Corresponding author.}, Bo Jia, Haonan Feng, Jingjing Guan, Yueshuang Jiao, Xiangdong Li \\
  Beijing University of Posts and Telecommunications \\
  \texttt{\{fengchen666, lihuill, q\_st, fenghaonan222, guaner, yueshuang, lxd\_dong\}@bupt.edu.cn} \\
}
\begin{document}
\flushbottom
\maketitle
\begin{abstract}
Formal verification provides rigorous guarantees for cryptographic security, yet extracting formalizable security goals from natural-language protocol documents remains largely manual. We introduce SecGoal, a dedicated expert-annotated dataset and benchmark for extracting formalizable security goal statements from protocol documents, covering 15 widely deployed protocols, together with AIFG, a schema- and flow-conditioned framework for structured formal security property generation. Our evaluation shows that frontier and large LLMs achieve high property recall but low extraction precision because they often fail to distinguish formalizable security goals from non-goal protocol content. In contrast, SecGoal fine-tuning makes smaller open-source LLMs substantially more selective extractors of formalizable security goals. On the held-out test protocols, Gemma2-9B-FT improves extraction precision from 24.0\% to 66.6\% and reaches 97.6\% property recall, outperforming larger prompted LLMs and encoder baselines. In a controlled setting, AIFG shows that concise goal inputs can support high-recall structured property generation, while expert-vetted extracted inputs reveal over-generation as the main remaining bottleneck. Together, SecGoal and AIFG provide a dataset, benchmark, and framework for specification-grounded security goal extraction and property generation.
\end{abstract}

\section{Introduction}
\begin{figure}[!t]
    \centering
    \vspace{-0.3cm}
    \includegraphics[width=\columnwidth]{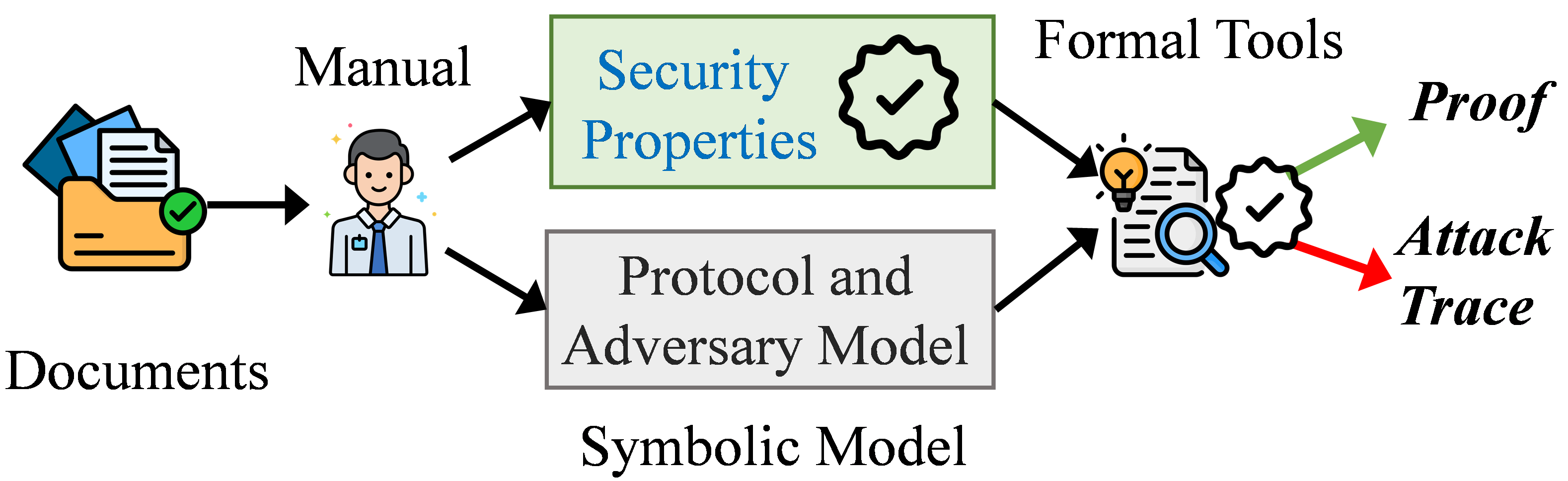}

 \vspace{-0.1cm}

    \caption{Protocol formal verification workflow from natural-language specification review to security goal identification and formal analysis.}
    \label{fig:fm}
        \vspace{-0.3cm}
\end{figure}

Formal verification is critical for cryptographic security, providing rigorous guarantees for widely deployed protocols such as TLS~1.3~\cite{bhargavan2017verified,cremers2017comprehensive} and 5G-AKA~\cite{basin2018formal,cremers2019component}.
As illustrated in Figure~\ref{fig:fm}, formal verification typically requires domain experts to manually curate documents and encode protocol logic, threat models, and security goals---such as confidentiality and authentication---into code for tools like Tamarin~\cite{meier2013tamarin} and ProVerif~\cite{blanchet2018proverif}. Upon execution, these tools can automatically generate attack traces for subtle flaws that evade conventional testing, or conversely, provide mathematical proofs that the protocol satisfies its intended security properties.
Crucially, the effectiveness of formal verification depends on an intermediate step that remains largely manual: extracting security goal statements from protocol documents and mapping them into precise symbolic security property descriptions.

This step is challenging because protocol documents, such as IETF RFCs, are written in long-form, unstructured natural language, while
verification tools operate on precise security properties. However, security goals are rarely stated explicitly; instead, they are dispersed across descriptive text, threat discussions, and design rationales. Even when identified, translating these goals into formal, tool-checkable property statements remains largely unsystematic, creating a persistent specification-to-model gap that limits the scalability of formal protocol analysis.

Recent advances in large language models (LLMs) have sparked interest in autoformalization---the automated translation of natural language into formal specifications~\cite{wu2022autoformalization}---and formal specification inference for programs~\cite{le2025can}. LLMs have demonstrated strong capabilities in code generation~\cite{chen2021evaluating}, mathematical proof synthesis~\cite{jiang2022draft}, and assertion generation~\cite{cosler2023nl2spec}, suggesting their potential to assist in automating security goal statement extraction and formalization.

However, security protocols pose a particularly difficult setting: documents often extend over hundreds of pages, embed goals implicitly, and require long-context semantic reasoning, where errors can lead to incomplete or unsound formal analyses. 
These observations raise a more focused question: \textit{can LLMs reliably extract formalizable security goals from long protocol documents, and to what extent can such extracted goals support structured formal security-property generation under a given protocol flow and formal schema?}

Existing work on LLMs for formal methods primarily focuses on code-level annotations or state-machine inference~\cite{wael2025agentic}, while publicly available datasets aligning natural-language protocol documents with formal security properties are absent. This data gap and lack of benchmarks prevents systematic evaluation of LLM capabilities in this high-stakes setting and hinders the development of LLM-based methods for automatic formal analysis of security protocols.

To address this gap, we formulate formalizable security-goal extraction as a benchmark task and introduce SecGoal, a dedicated expert-annotated dataset for this setting, together with AIFG, a schema- and flow-conditioned framework for structured formal security property generation from extracted goals. Our study asks whether LLMs can distinguish formalizable goals from surrounding protocol text, and how extraction quality affects downstream formalization. We find a clear gap between promise and practice: frontier LLMs such as GPT-5.4, Claude-Sonnet-4-6, and Gemini-3-Pro-Preview achieve high recall but lack the precision needed for formal analysis, while instruction tuning on SecGoal substantially improves extraction and enables accurate security property generation.

Our main contributions are as follows:
\begin{itemize} [leftmargin=*, nosep]
    \item We introduce SecGoal, an expert-annotated benchmark for extracting formalizable security goals from 15 widely deployed protocols, where goals are sparse and often implicitly embedded in real, heterogeneous specifications rather than always being stated explicitly.

    \item We define cross-granularity metrics that jointly measure extraction precision and security property coverage, and use them to systematically evaluate frontier and large LLMs, revealing a consistent high-recall, low-precision failure mode across the benchmark.

    \item We show that fine-tuning smaller open-source LLMs on SecGoal substantially improves this precision--recall trade-off, and provide AIFG, a schema- and flow-conditioned framework for security property generation from extracted goals.
\end{itemize}

\section{Related Work}

\paragraph{LLMs for Autoformalization.}
A growing body of research leverages LLMs to translate natural language constraints into formal artifacts for general software and hardware verification. Tools such as NL2Spec~\cite{cosler2023nl2spec} and recent approaches by Beg et al.~\cite{beg2025short} focus on generating code-level assertions (e.g., Dafny, JML) or function contracts~\cite{pearce2023examining}. Similarly, in hardware, AssertLLM~\cite{yan2025assertllm} automates the generation of SystemVerilog Assertions.
However, as highlighted by FormalBench~\cite{le2025can}, general-purpose models struggle with the logical rigor required for verification.
Crucially, these approaches operate in deterministic contexts (e.g., source code or localized comments). They do not address the linguistic challenge of extracting global security goals from unstructured, ambiguous protocol documents, where the ``code'' does not yet exist.

\paragraph{Automated Protocol Modeling.}
Prior work on protocol analysis typically assumes structured inputs or focuses on operational flows rather than security logic.
One stream translates semi-formal notations (e.g., Alice \& Bob) into verification models using compilers~\cite{keller2014converting} or LLMs~\cite{li2025constructing}.
Regarding operational extraction, RFCNLP~\cite{pacheco2022automated} and PROSPER~\cite{sharma2023prosper} utilize BERT-based architectures to extract Finite State Machines (FSMs). More recently, AutoSM~\cite{mao2025llm} leveraged LLMs and intermediate lambda calculus to synthesize executable symbolic models.
However, these methods face a dual limitation: they either require manual intermediate translation or overlook the extraction of formalizable security goals.

\paragraph{Datasets and Benchmarks.}
Existing resources largely bypass the extraction challenge. CryptoFormalEval~\cite{curaba2024cryptoformaleval} benchmarks the translation of pre-defined properties into Tamarin code, assuming the extraction is already done.
In the smart contract domain, PropertyGPT~\cite{liu2024propertygpt} relies on source code structure and compiler feedback, which are unavailable in natural language documents.

To our knowledge, SecGoal is the first expert-annotated benchmark specifically targeting the extraction of formalizable security goal evidence from protocol documents. Complementing this resource, AIFG serves as a schema- and flow-conditioned framework for structured security property generation from extracted goals.

\begin{figure*}[!t]
    \centering
    \vspace{-0.8cm}
    \includegraphics[width=1\textwidth]{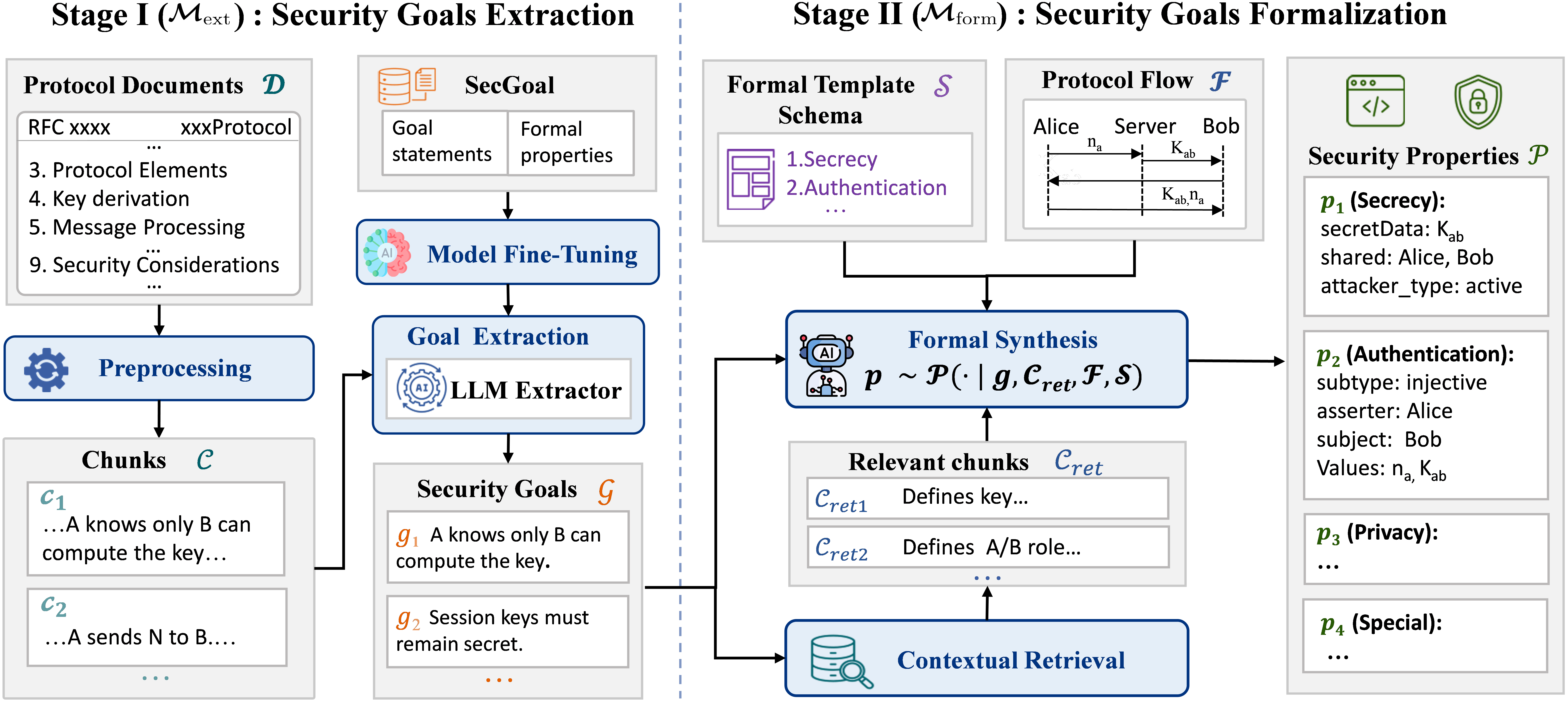}
    \caption{An overview of the AIFG framework.}
    \label{fig:AIFG}
    \vspace{0.05cm}
\end{figure*}
\section{The SecGoal Dataset}
\subsection{Protocol Coverage and Sources}
SecGoal covers 15 widely deployed protocols across critical domains---telecommunications (5G-AKA), industrial IoT (OPC-UA, SPDM), and web security (TLS 1.3, FIDO2)---whose specifications are drawn from heterogeneous sources, ranging from structured IETF RFCs to industrial standards and academic literature. Unlike synthetic datasets, SecGoal reflects real-world complexity: security goals are rarely stated explicitly but instead implicitly embedded in verbose, domain-specific prose.

\subsection{Human-in-the-Loop Annotation}
\label{sec:hitl}
We employed a three-phase HITL pipeline. First, an LLM with multiple queries generated broad candidate security goal statements to maximize coverage, shifting human effort from raw extraction to precise verification.

Second, two doctoral researchers specializing in cryptographic protocol analysis and formal verification independently reviewed the candidate security goals, removing unsupported candidates and adding omitted ones. We treated protocol specifications as the annotation ground truth, using formal-analysis literature only as a secondary reference. This is necessary because the two sources may differ in scope: \citet{basin2018formal} identify 5G-AKA goals and assumptions not explicit in the 3GPP specification, while some security goals stated in the EDHOC specification \citep{ietf-lake-edhoc-12} fall outside the properties explicitly analyzed by \citet{jacomme2023comprehensive}. Thus, formal analyses guided and validated our annotations but did not override the specifications.

Finally, a senior professor with domain expertise adjudicated disagreements and edge cases to ensure consistency across all 15 protocols.

Inter-annotator agreement is strong: statement selection achieves 0.844 F1 and 0.737 Jaccard similarity, while property labeling achieves 0.956 micro-F1 and 0.947 Cohen's $\kappa$. Per-protocol results are provided in Appendix~\ref{app:iaa}.
\subsection{Dataset Statistics and Split}

SecGoal contains 391 annotated security goal statements and 131 formal security
properties across 15 protocols. These statements occupy only 2.5\% of the source
text on average, creating a sparse and challenging extraction setting.
Table~\ref{tab:dataset_stats} in the appendix reports per-protocol statistics.
For evaluation, we use protocol-level splits,
ensuring that all text from a protocol appears in only one split.
\section{AIFG Framework}
AIFG is a schema- and flow-conditioned framework for structured security property generation. It extracts formalizable security goal statements and generates structured property descriptions using retrieved context, an available protocol flow, and a formal schema, but does not synthesize executable formalization tool code from raw documents.
\subsection{Overview}
\label{sec:overview}


AIFG bridges protocol documents and formal security properties by extracting
security goals from $\mathcal{D}$ and mapping them, conditioned on a selected
protocol flow $\mathcal{F}$ and schema $\mathcal{S}$, to a set of canonical
formal property descriptions $\mathcal{P} = \{p_1, \dots, p_m\}$.

Rather than attempting end-to-end translation, AIFG decomposes the pipeline into two stages: extraction and formalization. As illustrated in Figure~\ref{fig:AIFG}, Stage I ($\mathcal{M}_{\mathrm{ext}}$) extracts security goal statements $\mathcal{G}$ from $\mathcal{D}$, filtering out operational content. Stage II ($\mathcal{M}_{\mathrm{form}}$) performs retrieval-augmented formalization: it first uses each extracted goal $g \in \mathcal{G}$ to retrieve relevant contextual chunks $\mathcal{C}_{\mathrm{ret}}$, and then synthesizes structured formal properties $\mathcal{P}$ conditioned on $\mathcal{C}_{\mathrm{ret}}$, the protocol flow $\mathcal{F}$, and the formal template schema $\mathcal{S}$.

\subsection{Stage I: Security Goal Extraction}
Security-goal statements are sparse in protocol documents, covering only
2.5\% of annotated document text on average. Directly mapping a full document
to formal properties is prone to context dilution, as operational descriptions
dominate the input while formalizable goals remain rare. Stage I therefore uses
the extractor $\mathcal{M}_{\mathrm{ext}}$ to extract security goal statements from
$\mathcal{D}$ and produce the goal set $\mathcal{G}$ for security
formalization.

\paragraph{Document Preprocessing.}
We first remove structural boilerplate, such as tables of contents,
bibliographies, and copyright notices, to reduce irrelevant noise. The remaining
document text is then segmented to mitigate context window
limitations~\cite{liu2024lost} and retain sufficient local context for
extraction. 
This preprocessing step yields a sequence of chunks
$\mathcal{C}=\{c_1,\dots,c_n\}$, which serve as input to the extractor.


\paragraph{Model Fine-Tuning.}

AIFG is model-agnostic, but we instantiate $\mathcal{M}_{\mathrm{ext}}$ with a model
fine-tuned on SecGoal to improve extraction accuracy. Since security goal
statements are rare, the training data is highly imbalanced and can bias the
model toward non-goal protocol content. We therefore adopt Randomized Negative
Downsampling~\cite{liu2008exploratory}, sampling negative examples to maintain a
controlled positive-to-negative ratio (e.g., 1:3) and improve discrimination
between formalizable goals and non-goal text.



\paragraph{Goal Extraction.}
During extraction, the extractor $\mathcal{M}_{\mathrm{ext}}$ is applied to each chunk
$c_i \in \mathcal{C}$. If no security goal statement is detected, the extractor
returns an empty list; otherwise, it generates a structured list of the
formalizable security goal statements in $c_i$. The extracted statements are
then aggregated and deduplicated to form the protocol-level candidate goal set
$\mathcal{G}$.

\subsection{Stage II: Security Goal Formalization}

Translating the security goal statements $\mathcal{G}$ into formal
security properties that are accurate and consistent with the formal protocol
model is non-trivial, due to several linguistic and logical challenges:
\begin{itemize}[leftmargin=*, nosep]
\item \textbf{Ambiguity:} Narrative goals often omit explicit role references or agreement variables,
such as which nonces or transcript hashes constitute the authentication
\texttt{agreementValues}. These details must be grounded using the protocol context.
\item \textbf{Symbolic mismatch:} Natural-language terms do not necessarily
align with the variable names, role names, and message fields used in the target
formal model.
\item \textbf{Mapping complexity:} The mapping is many-to-many: a single goal
statement may imply multiple formal properties, while multiple goal statements
may support the same property.
\end{itemize}

\paragraph{Contextual Retrieval.}
Because entity definitions and variable bindings are scattered across
$\mathcal{D}$, translating each goal in isolation can cause symbolic mismatches.
For each extracted goal $g \in \mathcal{G}$, we retrieve the top-$k$ relevant
chunks $\mathcal{C}_{\mathrm{ret}}$ using $g$ as a semantic query. The retrieved context
helps ground underspecified roles, entities, and protocol terms, and is combined
with the protocol variable structure described next.

\paragraph{Flow and Schema Inputs.}
AIFG treats the protocol flow $\mathcal{F}$ as a model-conditioned input rather
than an artifact automatically inferred from raw documents, because symbolic
protocol models may vary across experts and configurations. Here,
$\mathcal{F}$ is a structured representation of a selected complete formal
protocol-flow model, including the roles, ordered message exchanges, and
flow-level variables used to align document terms with formal-model notation.
The schema $\mathcal{S}$ is a protocol-independent structured representation of
formal security properties. It summarizes common property specifications into
typed slots, providing a canonical space for property generation, evaluation,
and many-to-many deduplication. The full template definitions are provided in
Appendix~\ref{sec:appendix_templates}.
\paragraph{Formal Property Synthesis.}
To ground retrieved goals in the target formal model, we combine
$\mathcal{C}_{\mathrm{ret}}$ with $\mathcal{F}$ and $\mathcal{S}$. For each goal
$g \in \mathcal{G}$, the model generates each candidate property $p$ according
to $p \sim\mathcal{P}(\cdot \mid g, \mathcal{C}_{\mathrm{ret}}, \mathcal{F}, \mathcal{S})$.
Guided by $\mathcal{S}$, the model instantiates structured security property templates
and maps natural-language terms to canonical variables in $\mathcal{F}$; for
example, a phrase such as ``shared secret'' may be grounded to $K_{AB}$ and
placed in the \texttt{secretData} slot. Finally, all generated properties are
canonicalized and deduplicated to obtain the final security property set $\mathcal{P}$.
\section{Evaluation}
We evaluate LLMs and AIFG on SecGoal, guided by
three research questions:
\begin{itemize}[leftmargin=*, nosep]
    \item \textbf{RQ1:} \textit{What limitations do contemporary LLMs exhibit
    when extracting security goals from protocol documents?}
    \item \textbf{RQ2:} \textit{How effectively does SecGoal fine-tuning improve
    security goal extraction, and can fine-tuned small generative models
    outperform encoder baselines and frontier LLMs?}
    \item \textbf{RQ3:} \textit{How reliably can AIFG generate structured
    formal security-property descriptions from extracted security goals?}
\end{itemize}

We use LlamaFactory~\cite{zheng2024llamafactory} for the RQ2 fine-tuning
experiments and RAGFlow~\cite{ragflow2024} for the RQ3 formalization pipeline.
For RQ1 and RQ2, all LLMs are evaluated with provider-recommended default
settings, with temperature fixed to 0 for determinism and reproducibility.
Detailed configurations for all experiments are provided in
Appendix~\ref{sec:appendix_setup}. The formal template schema and prompt
templates are presented in Appendices~\ref{sec:appendix_templates}
and~\ref{sec:appendix_prompts}, respectively.

\begin{figure*}[!t]
    \vspace{-0.9cm}
    \centering
    \includegraphics[width=\textwidth]{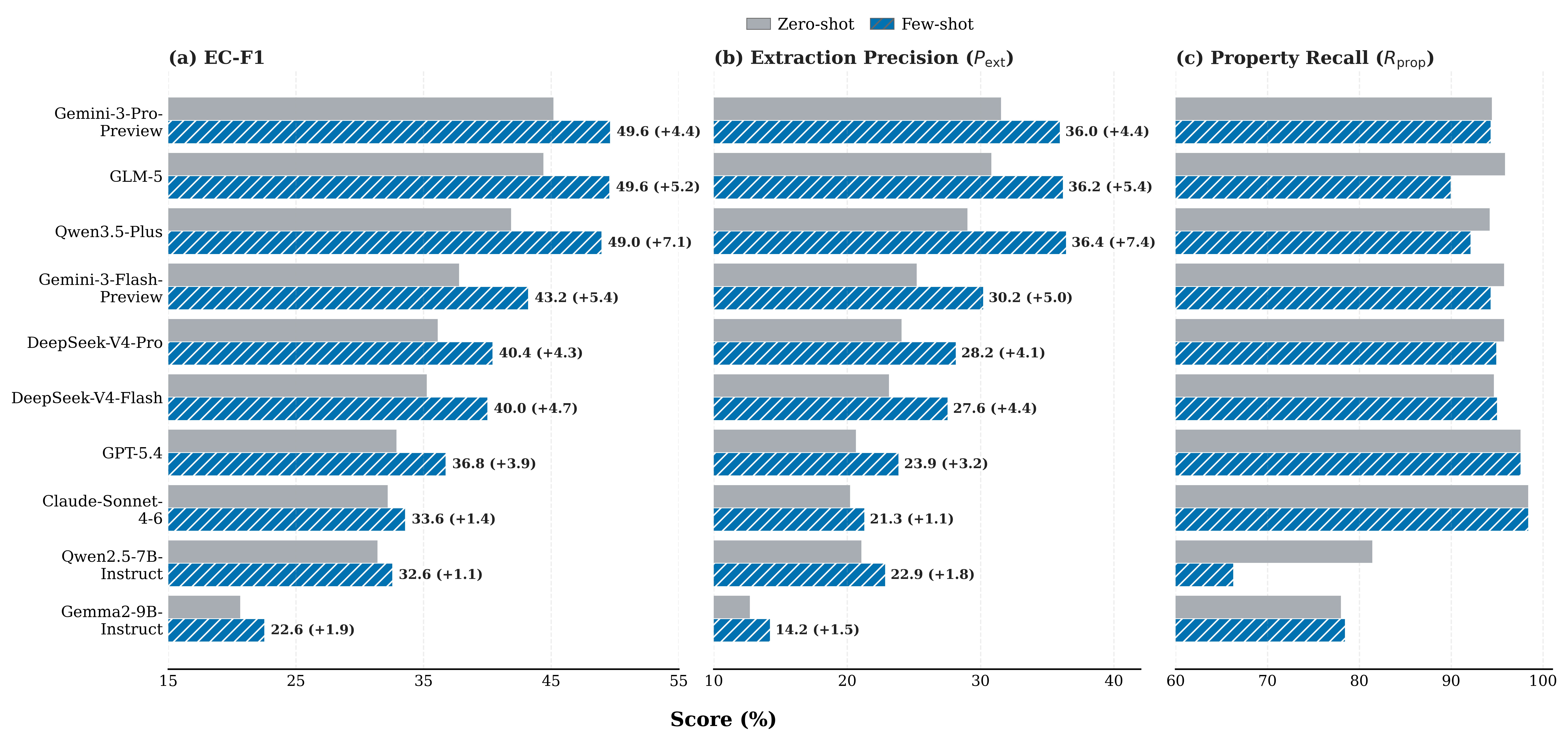}
    \vspace{-0.6cm}
    \caption{Zero-shot and few-shot extraction performance across models, averaged over all 15 SecGoal protocols. Models are ordered by few-shot EC-F1. Few-shot prompting improves precision substantially while property recall remains largely saturated, yielding higher EC-F1 overall.}
    \label{fig:rq1_three_panel}
\end{figure*}

\subsection{Evaluation Metrics}

We employ a specialized metric suite that extends PropertyGPT~\cite{liu2024propertygpt}
to address the non-bijective mapping inherent to our task: a single textual
statement may encode multiple properties, and one property may be supported by several
statements, rendering standard statement-level or property-level metrics
insufficient alone for this task. Detailed definitions and a worked example are provided in
Appendix~\ref{sec:appendix_metrics}.

\paragraph{Metrics for Extraction (RQ1 \& RQ2).}
We jointly evaluate the precision of extracted security goal statements and
property-level coverage via three complementary metrics:

\begin{itemize}[leftmargin=*, noitemsep]

    \item \textbf{Extraction Precision ($P_{\mathrm{ext}}$):}
    $P_{\mathrm{ext}} = \frac{TP}{TP+FP}$, where $TP$ counts extracted statements
    that are valid security goal statements and $FP$ counts extracted statements
    that are not, measuring the signal-to-noise ratio of the extraction.

    \item \textbf{Property-level Recall ($R_{\mathrm{prop}}$):}
    $R_{\mathrm{prop}} = \frac{|\mathcal{P}_{hit}|}{|\mathcal{P}_{GT}|}$,
    where $\mathcal{P}_{hit}$ denotes the set of ground-truth formal properties
    whose evidential statements are present in the extracted output, and
    $\mathcal{P}_{GT}$ is the complete ground-truth property set.
    This measures the fraction of formal properties recoverable downstream.

    \item \textbf{Extraction-Coverage F1 (EC-F1):}
    $\mathrm{EC\text{-}F1} = \frac{2 \cdot P_{\mathrm{ext}} \cdot R_{\mathrm{prop}}}
    {P_{\mathrm{ext}} + R_{\mathrm{prop}}}$, a cross-granularity harmonic mean
    grounded in the requirements of formal verification: $P_{\mathrm{ext}}$
    minimizes the noise that propagates into the formalization pipeline, while
    $R_{\mathrm{prop}}$ ensures no critical security properties are missed during
    formal analysis. We adopt EC-F1 as the primary extraction metric.

\end{itemize}





\paragraph{Metrics for Formalization (RQ3).}
We assess formal property generation along two complementary dimensions:

\begin{itemize}[leftmargin=*, noitemsep]

    \item \textbf{Property-level Precision/Recall/F1:}
    Measures whether the generated property set recovers the gold formal
    security properties. A generated property is counted as a match only when it
    has the same property type and subtype as a gold property and is sufficiently
    aligned in its roles and core security data. Precision penalizes spurious
    properties, while recall measures how many gold properties are recovered.

    \item \textbf{Slot-level Precision/Recall/F1:}
    Measures whether the internal fields of matched properties are grounded
    correctly. We flatten each property into slot tokens such as
    \texttt{asserter=R} or \texttt{agreementValues=TH\_2}, then compare the
    generated and gold slot-token sets. This captures errors in roles,
    attacker assumptions, keys, nonces, transcript hashes, and other formal
    arguments even when the property type is correct. Slot scores are computed
    over matched property pairs; unmatched generated and gold properties are
    reflected by property-level precision and recall. Full details are provided
    in Appendix~\ref{sec:appendix_metrics}.

\end{itemize}

\subsection{RQ1: LLM Performance on Goal Extraction}
To answer RQ1, we evaluate several frontier and large open-source LLMs, together
with two compact instruction-tuned models, under zero-shot and few-shot
prompting. Figure~\ref{fig:rq1_three_panel} reports property-level recall
($R_{\mathrm{prop}}$), extraction precision ($P_{\mathrm{ext}}$), and EC-F1
averaged over all 15 SecGoal protocols.

\paragraph{Off-the-shelf LLMs achieve high recall but low precision.}
Among frontier and large prompted LLMs, $R_{\mathrm{prop}}$ is consistently high: zero-shot recall ranges from 94.24\% to 98.43\%, indicating that these models can usually locate regions containing recoverable security goals. The compact instruction-tuned baselines are less recall-stable, but the same precision bottleneck remains across the benchmark. Zero-shot $P_{\mathrm{ext}}$ ranges only from 20.25\% to 31.57\% among the stronger prompted models, and even few-shot prompting raises the best precision only to 36.44\% for Qwen3.5-Plus. Consequently, no prompted model reaches 50\% EC-F1; the strongest few-shot results are 49.65\% for Gemini-3-Pro-Preview, 49.59\% for GLM-5, and 48.97\% for Qwen3.5-Plus.

This gap reveals the central failure mode of current LLMs: they are good at detecting broadly security-related content, but poor at separating formalizable security goals from protocol mechanisms, implementation constraints, and explanatory text. For formal verification, such low precision is costly, because each spurious extraction must be manually filtered before formalization.


\paragraph{Few-shot prompting only partially mitigates the gap.}
Few-shot prompting improves $P_{\mathrm{ext}}$ and EC-F1 for several models,
with the largest EC-F1 gain observed for Qwen3.5-Plus (+7.08 points). However,
the improvement remains limited and sometimes trades recall for precision:
GLM-5 decreases from 95.90\% to 89.99\% in $R_{\mathrm{prop}}$, and
Qwen2.5-7B-Instruct drops from 81.45\% to 66.31\%.

Model scale alone also does not guarantee better extraction. GPT-5.4 and
Claude-Sonnet-4-6 achieve very high few-shot recall, 97.60\% and 98.43\%,
respectively, but their precision remains only 23.88\% and 21.31\%.
Conversely, Qwen3.5-Plus obtains the highest few-shot precision, 36.44\%, but
still trails Gemini-3-Pro-Preview and GLM-5 in EC-F1. Overall, the best EC-F1
scores come from models with a better precision--recall balance rather than
from models maximizing either dimension alone.

\paragraph{Answer to RQ1.}
Contemporary LLMs are not reliable standalone extractors for security protocol formalization. Frontier and large prompted models often achieve high property coverage, but all prompted models systematically over-extract non-goal text. Few-shot prompting improves selectivity only marginally, indicating that domain-specific adaptation is necessary.

\begin{figure*}[!t]
    \centering
    \vspace{-0.25cm}
    \begin{minipage}[t]{0.48\textwidth}
        \centering
        \includegraphics[height=0.185\textheight,keepaspectratio]{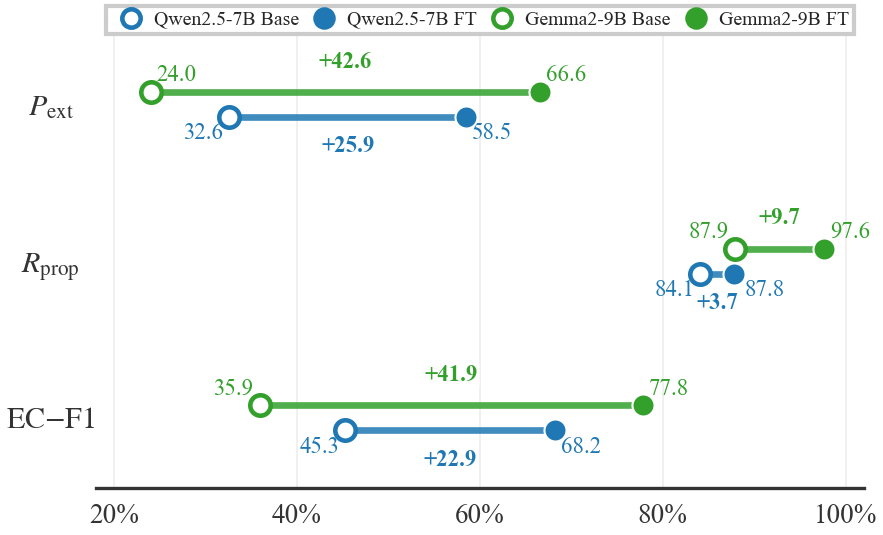}
        \vspace{-0.3cm}
        \makebox[\linewidth][c]{\footnotesize (a) Instruct vs.\ fine-tuned models}
    \end{minipage}\hfill
    \begin{minipage}[t]{0.48\textwidth}
        \centering
        \includegraphics[height=0.205\textheight,keepaspectratio]{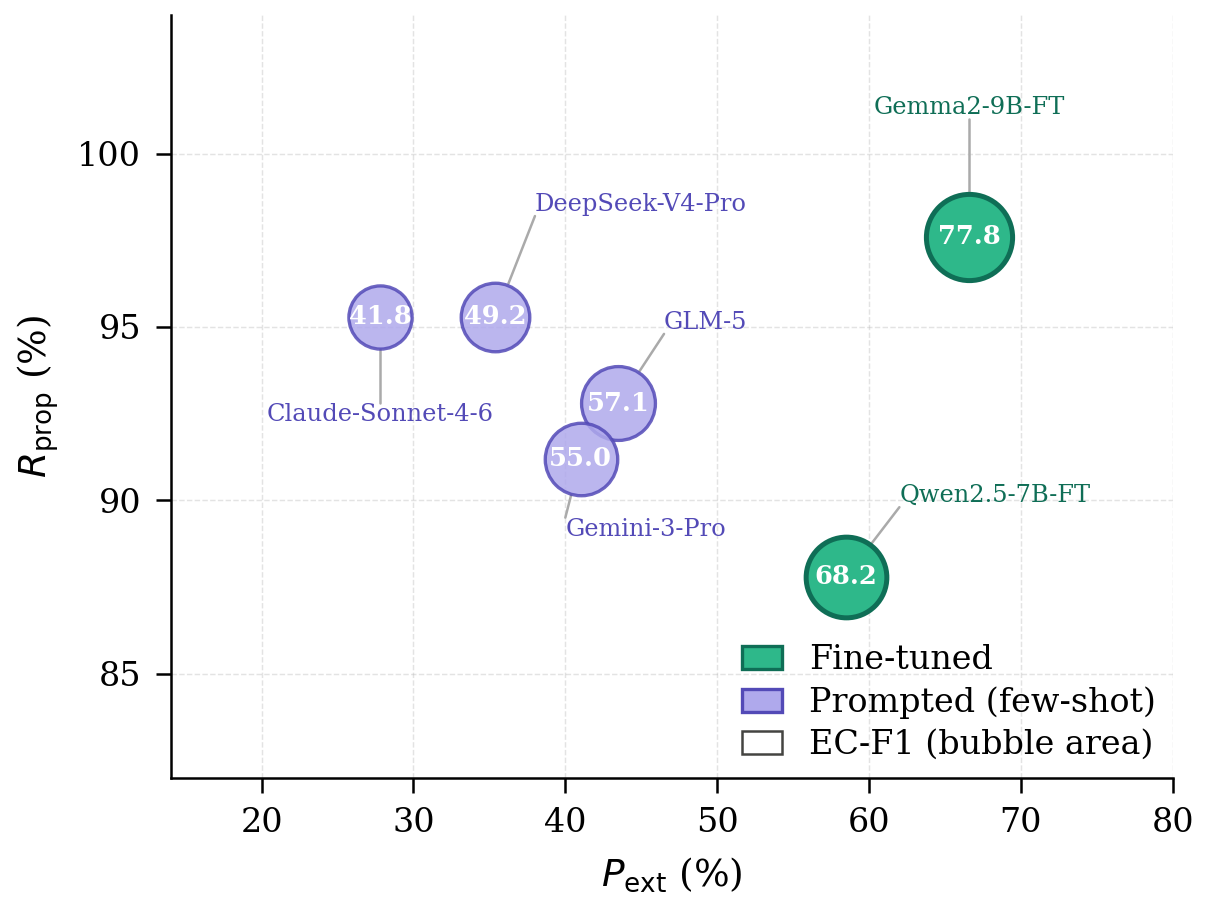}
        \vspace{-0.3cm}
        \makebox[\linewidth][c]{\footnotesize (b) Fine-tuned models vs.\ prompted LLMs}
    \end{minipage}
    \caption{RQ2 held-out test results: (a) instruction-tuned models versus SecGoal fine-tuned models; (b) fine-tuned small models versus few-shot LLMs, with bubble area denoting EC-F1.}
    \label{fig:rq2-ft}
    \label{fig:rq2-2}
    \vspace{0.3cm}
\end{figure*}

\subsection{RQ2: Effectiveness of SecGoal}
To answer RQ2, we evaluate whether SecGoal fine-tuning improves smaller
open-source LLMs as security goal extractors. We compare Gemma2-9B and
Qwen2.5-7B before and after fine-tuning, and benchmark the fine-tuned models
against encoder baselines and strong few-shot prompted LLMs. All RQ2 results
are computed on the held-out test protocols after training on the SecGoal training
split. Accordingly, the prompted baselines in Figure~\ref{fig:rq2-2}(b) are
re-evaluated on the same test subset rather than using the all-protocol averages
from RQ1.

\paragraph{Fine-tuning corrects the precision--recall imbalance.}
Figure~\ref{fig:rq2-ft}(a) shows that SecGoal fine-tuning improves both smaller open-source LLMs. For Gemma2-9B, $P_{\mathrm{ext}}$ rises from 24.0\% to 66.6\% on the held-out test set, while $R_{\mathrm{prop}}$ increases from 87.9\% to 97.6\%, raising EC-F1 from 35.9\% to 77.8\%. Qwen2.5-7B shows the same pattern: $P_{\mathrm{ext}}$ improves from 32.6\% to 58.5\%, $R_{\mathrm{prop}}$ from 84.1\% to 87.8\%, and EC-F1 from 45.3\% to 68.2\%.

These gains suggest that SecGoal fine-tuning primarily helps models draw a
clearer and more reliable boundary between formalizable security goals and
related but non-goal protocol text. While prompted LLMs often retrieve broadly
relevant passages, fine-tuning provides the domain-specific supervision needed
to filter out operational descriptions that should not be formalized, yielding
more selective and trustworthy extraction behavior.

\paragraph{Context and negative-example downsampling both matter.}
Table~\ref{tab:ablation} shows that section-level context and randomized
negative-example downsampling are complementary.
Fixing the chunking strategy at section-level, downsampling improves EC-F1
from 70.5\% (no downsampling) to 77.8\% (1:3 ratio);
fixing the downsampling ratio at 1:3, section-level chunking outperforms
sentence-window chunking by 3.2 points (77.8\% vs.\ 74.6\%).
This indicates that effective extraction requires both contextual continuity
and balanced exposure to goal and non-goal examples.

\paragraph{Fine-tuned compact LLMs outperform encoder and prompted LLM baselines.}
Table~\ref{tab:encoder_baselines} shows that Gemma2-9B-FT outperforms encoder
baselines trained on the same split, achieving 77.78\% EC-F1 versus
56.21\% for RoBERTa~\cite{liu2019roberta} and 66.32\% for
SecureBERT~\cite{aghaei2022securebert}. The gain over SecureBERT is mainly
recall-driven, while precision also improves modestly.

The fine-tuned LLMs also surpass much larger prompted models in
Figure~\ref{fig:rq2-2}(b). Gemma2-9B-FT exceeds the strongest prompted baseline,
GLM-5, by 20.7 EC-F1 points, while Qwen2.5-7B-FT also outperforms all prompted
baselines. The main advantage is precision: Gemma2-9B-FT reaches 66.6\%
$P_{\mathrm{ext}}$, whereas prompted models remain below 45\% on the same test split.

\paragraph{Answer to RQ2.}
SecGoal fine-tuning improves extraction by teaching models to distinguish
formalizable security goals from non-goal protocol text. Compact fine-tuned LLMs
outperform both encoder specialists and larger prompted LLMs, indicating that
task-specific supervision matters more than scale alone for this task.

\begin{table}[!tbp]
\centering
\footnotesize
\vspace{-0.2cm}
\begingroup
\setlength{\tabcolsep}{3pt}
\renewcommand{\arraystretch}{0.96}
\begin{tabular}{ll ccc}
\toprule
\textbf{Chunking} & \textbf{Downsamp.} & $P_{\mathrm{ext}}$ & $R_{\mathrm{prop}}$ & EC-F1 \\
\midrule
\multirow{2}{*}{Sentence-window}
  & None   & 63.1 & 83.4 & 70.3 \\
  & 1{:}3  & 65.7 & 90.4 & 74.6 \\
\midrule
\multirow{2}{*}{Section-level}
  & None   & 61.2 & 86.0 & 70.5 \\
  & 1{:}3  & \textbf{66.6} & \textbf{97.6} & \textbf{77.8} \\
\bottomrule
\end{tabular}
\endgroup
\vspace{-0.2cm}
\caption{Ablation results on held-out test protocols with Gemma2-9B-FT\@ under the same train/test split.}
\label{tab:ablation}
\vspace{0.05cm}
\end{table}
\begin{table}[!tbp]
\centering
\small
\setlength{\tabcolsep}{4pt}
\renewcommand{\arraystretch}{1.08}
\begin{tabular}{llccc}
\toprule
\textbf{Family} & \textbf{Model} & $R_{\mathrm{prop}}$ & $P_{\mathrm{ext}}$ & \textbf{EC-F1} \\
\midrule
Encoder & RoBERTa & 62.25 & 57.91 & 56.21 \\
Encoder & SecureBERT & 79.75 & 59.97 & 66.32 \\
LLM & Gemma2-9B-FT & \best{97.65} & \best{66.61} & \best{77.78} \\
\bottomrule
\end{tabular}
\vspace{-0.2cm}
\caption{Encoder results on held-out test protocols using the same train/test split as the fine-tuned LLMs.}
\label{tab:encoder_baselines}
\vspace{-0.1cm}
\end{table}


\subsection{RQ3: Formal Property Generation}

To answer RQ3, we evaluate AIFG as a flow-conditioned property-generation task:
given extracted goals, a protocol flow, and a schema, AIFG generates canonical
property descriptions and grounds their slots. We instantiate AIFG with GPT-5.4~\cite{openai2026gpt54}
and text-embedding-large~\cite{openai2024textembedding3large} retrieval; details are in
Appendix~\ref{sec:appendix_setup}. For each protocol, the flow $\mathcal{F}$ is
derived from formal-analysis papers and implementations and normalized
into AIFG's flow format; gold  properties are written relative to this
flow and schema $\mathcal{S}$. We test four protocols spanning compact
academic and industrial settings (NSSK, Kao-Chow v1, EDHOC, and SPDM) under two input
variants: \textit{gold-minimal}, concise gold statements covering all gold
properties, and \textit{expert-vetted}, reviewed Gemma2-9B-FT extractions with
false positives removed. This setting evaluates intermediate structured property
generation, not end-to-end formal verification or executable model synthesis.

\begin{table}[!tbp]
    \vspace{-0.2cm}
\centering
\footnotesize
\setlength{\tabcolsep}{2pt}
\renewcommand{\arraystretch}{1.0}
\begin{tabular}{@{}lllccc@{}}
\toprule
\textbf{Protocol} & \textbf{Input} & \textbf{Level} & \textbf{P} & \textbf{R} & \textbf{F1} \\
\midrule
NSSK & gold-min. & Prop. & 0.71$_{\pm 0.00}$ & 1.00$_{\pm 0.00}$ & 0.83$_{\pm 0.00}$ \\
 &  & Slot & 0.93$_{\pm 0.09}$ & 1.00$_{\pm 0.00}$ & 0.96$_{\pm 0.05}$ \\
NSSK & expert-vet. & Prop. & 0.58$_{\pm 0.06}$ & 1.00$_{\pm 0.00}$ & 0.74$_{\pm 0.05}$ \\
 &  & Slot & 1.00$_{\pm 0.00}$ & 1.00$_{\pm 0.00}$ & 1.00$_{\pm 0.00}$ \\
\midrule
KC-v1 & gold-min. & Prop. & 0.73$_{\pm 0.19}$ & 0.73$_{\pm 0.19}$ & 0.73$_{\pm 0.19}$ \\
 &  & Slot & 0.89$_{\pm 0.08}$ & 1.00$_{\pm 0.00}$ & 0.94$_{\pm 0.04}$ \\
KC-v1 & expert-vet. & Prop. & 0.38$_{\pm 0.03}$ & 1.00$_{\pm 0.00}$ & 0.55$_{\pm 0.03}$ \\
 &  & Slot & 1.00$_{\pm 0.00}$ & 1.00$_{\pm 0.00}$ & 1.00$_{\pm 0.00}$ \\
\midrule
EDHOC & gold-min. & Prop. & 0.73$_{\pm 0.02}$ & 0.86$_{\pm 0.03}$ & 0.79$_{\pm 0.03}$ \\
 &  & Slot & 0.85$_{\pm 0.00}$ & 0.89$_{\pm 0.00}$ & 0.87$_{\pm 0.00}$ \\
EDHOC & expert-vet. & Prop. & 0.38$_{\pm 0.03}$ & 0.76$_{\pm 0.00}$ & 0.51$_{\pm 0.02}$ \\
 &  & Slot & 0.83$_{\pm 0.01}$ & 0.89$_{\pm 0.00}$ & 0.85$_{\pm 0.01}$ \\
\midrule
SPDM & gold-min. & Prop. & 0.55$_{\pm 0.08}$ & 0.92$_{\pm 0.12}$ & 0.69$_{\pm 0.09}$ \\
 &  & Slot & 0.77$_{\pm 0.08}$ & 0.98$_{\pm 0.02}$ & 0.84$_{\pm 0.06}$ \\
SPDM & expert-vet. & Prop. & 0.27$_{\pm 0.03}$ & 0.71$_{\pm 0.06}$ & 0.39$_{\pm 0.04}$ \\
 &  & Slot & 0.73$_{\pm 0.06}$ & 0.95$_{\pm 0.01}$ & 0.81$_{\pm 0.05}$ \\
\bottomrule
\end{tabular}
\vspace{-0.2cm}
\caption{RQ3 property- and slot-level formalization results (mean $\pm$ std., three runs); full TP/FP/FN counts are in Appendix Table~\ref{tab:rq3_complete_formalization}. KC-v1: Kao-Chow v1; gold-min./expert-vet.: gold-minimal/expert-vetted.}
\vspace{-0.15cm}
\label{tab:rq3_formalization}
\end{table}

\paragraph{Concise inputs improve coverage but not exact selection.}
Gold-minimal inputs give AIFG concise statements that cover the target property
set, improving property recall on most protocols (Table~\ref{tab:rq3_formalization}).
NSSK reaches full recall with 0.83 property F1, while EDHOC and SPDM reach
0.86 and 0.92 recall, respectively. However, property precision remains limited
by extra generated properties and strict canonical matching, yielding 0.79 F1
on EDHOC and 0.69 on SPDM; Kao-Chow v1 reaches 0.73 F1 despite strong slot
grounding, reflecting property-level matching sensitivity.

\paragraph{Expert-vetted inputs expose over-generation and coverage gaps.}
Under expert-vetted inputs, AIFG often retains useful coverage but generates a
larger property set. NSSK and Kao-Chow v1 recover all gold properties, yet
over-generation lowers property F1 to 0.74 and 0.55, respectively. EDHOC and
SPDM are harder: recall falls to 0.76 and 0.71, while additional generated
properties reduce F1 to 0.51 and 0.39. This contrast suggests a complexity
effect. For compact protocols, the flow and schema expose fewer canonical
distinctions, so AIFG mainly needs to suppress redundant properties after
coverage is achieved. By contrast, EDHOC and SPDM involve variable-rich
properties spanning state, session binding, negotiation, measurements, and
authentication. Their errors therefore combine
over-generation, missed canonical properties, and incomplete variable
grounding, which lowers slot F1 on matched pairs (0.85 and 0.81 under
expert-vetted inputs) relative to NSSK and Kao-Chow v1 (1.00). EDHOC further
shows an upstream coverage effect: its expert-vetted goals cover fewer gold
properties, reducing generated recall; the gold-minimal rows also show that
complete goal coverage does not guarantee complete structured property
generation.
\paragraph{Answer to RQ3.}
With concise, well-covered inputs, AIFG can recover many target properties and
ground matched slots reliably. Its main remaining bottleneck is selecting the
minimal canonical property set: realistic extractions amplify over-generation,
and complex protocols such as EDHOC and SPDM also expose coverage gaps.



\section{Conclusion}

We introduced SecGoal, a dedicated expert-annotated benchmark for security
goal extraction from protocol documents, and AIFG, a schema- and
flow-conditioned framework for generating structured formal security
properties from extracted goals. Our evaluation shows that frontier and large LLMs can often identify security-relevant passages but struggle to filter formalizable goals, making precision the main bottleneck. SecGoal fine-tuning substantially improves this boundary judgment, enabling smaller open-source LLMs to outperform larger prompted models and encoder baselines. 
AIFG further demonstrates that concise goal inputs support high-recall structured property generation, while expert-vetted extracted inputs expose over-generation as the key obstacle to scaling.

\section{Limitations}
\paragraph{Specification-grounded scope.}
SecGoal and AIFG extract and formalize security goals that are stated or
supported by the protocol specification. They do not synthesize additional
security assumptions or expert-derived properties that may be required for a
complete security analysis~\cite{basin2018formal}. The output should therefore
be viewed as a specification-grounded goal set, not as a proof of full protocol
security.

\paragraph{Gap to end-to-end verification.}
AIFG focuses on the goal side of formal verification: extracting security goals
and generating structured formal property descriptions. It assumes an available
protocol flow and variable namespace, and does not automatically synthesize
executable Tamarin or ProVerif models from raw documents. Although these tools
can automate reasoning over a specified model, completing a proof often requires
expert-written auxiliary lemmas, suitable proof strategies, and iterative model
or query refinements. Thus, fully automating the path from raw protocol
documents to completed formal verification remains an open challenge in both
research and practice.

\paragraph{Limited formalization scale.}
Our RQ3 evaluation covers NSSK, Kao-Chow v1, EDHOC, and SPDM, extending the
controlled setting from compact academic protocols to a more realistic industrial
protocol. Even so, the evaluation remains limited by the need to construct
expert-validated flows, variable namespaces, and gold structured properties with
high confidence. We do not include OPC-UA in RQ3 because its security behavior is
spread across multiple specifications and deployment profiles; even before
threat-model dimensions, OPC-UA yields 216 protocol configurations, making a
single gold flow difficult to define without introducing strong manual modeling
assumptions. Including it would make the evaluation depend heavily on the
completeness of an OPC-UA-specific model rather than isolating AIFG's
property-generation ability. We therefore treat RQ3 as a controlled
property-generation study rather than a coverage study over all SecGoal
protocols.

\section*{Ethical Considerations}

\noindent\textbf{Data and privacy.}
SecGoal is constructed from publicly available protocol specifications,
standards, and academic materials. It does not contain personal, sensitive, or
user-generated private data, and focuses on protocol-level security
requirements rather than information about individual users or organizations.
When releasing the dataset and code, we will respect the redistribution terms
of the original source documents and provide documentation of protocol
coverage, annotation criteria, and intended use.

\noindent\textbf{Annotation process.}
All annotation and adjudication were conducted internally by members of the
research team. Two doctoral researchers with experience in cryptographic
protocol analysis and symbolic verification independently reviewed the
candidate security goals, while a senior domain expert, also a member of the
research team, adjudicated disagreements and edge cases. No crowdworkers,
external paid annotators, or vulnerable populations were involved. The task
required expert judgments about specification-grounded, formalizable security
goals, rather than subjective or personal information from annotators.

\noindent\textbf{Security impact and dual use.}
Because this work concerns security protocols, we acknowledge potential
dual-use and over-reliance risks. The proposed methods may help analysts
identify formalizable security goals more efficiently, but their outputs should
not be interpreted as a proof of protocol security. SecGoal and AIFG are
specification-grounded: they extract and formalize goals that are stated in or
supported by the source documents, but they do not synthesize all
expert-derived assumptions needed for complete security analysis.

\noindent\textbf{Intended use.}
This benchmark is intended to support research on specification-grounded
security goal extraction and structured security-property generation. We
discourage using the system as a standalone security certification tool, as a
substitute for expert formal verification, or as the sole basis for
security-critical deployment decisions. Outputs produced by models trained or
evaluated on SecGoal should be reviewed by qualified domain experts before
being used in formal analysis workflows or other security-sensitive settings.

\bibliography{custom}
\clearpage

\appendix
\label{sec:appendix}
\section{Experimental Setup and Hyperparameters}
\label{sec:appendix_setup}
\subsection{Hardware and Software}

All local fine-tuning experiments were conducted on a server equipped with
1 $\times$ NVIDIA A800 80GB PCIe GPU, an Intel Xeon Gold 6348 CPU with
112 logical cores, and 1.0 TiB system memory. The operating system was
Ubuntu 22.04.5 LTS\@. The NVIDIA driver version was 580.82.07, and the CUDA
runtime version reported by \texttt{nvidia-smi} was 13.0. We used
LlamaFactory~\cite{zheng2024llamafactory} for fine-tuning the open-source
LLM extractors.

Prompting-based evaluations of proprietary LLMs were conducted through their
official hosted interfaces or APIs. To improve reproducibility, we fixed the
temperature to 0 for all prompting experiments, while keeping all other
inference parameters at the officially recommended default settings. For models
that provide optional thinking or extended-reasoning modes, we disabled these
modes in all experiments and used standard prompted inference. The
hardware configuration above applies only to local fine-tuning experiments.
\subsection{Fine-tuning and Hyperparameters}

We fine-tuned Qwen2.5-7B-Instruct and Gemma2-9B-Instruct using LoRA-based
supervised fine-tuning in LlamaFactory~\cite{zheng2024llamafactory}. For each
model, we used the same fine-tuning configuration across its corresponding
experimental variants, while changing only the training data construction when
needed for ablation analysis.

\begin{table}[ht]
\centering
\footnotesize
\setlength{\tabcolsep}{3pt}
\renewcommand{\arraystretch}{1.08}
\begin{tabularx}{\columnwidth}{@{}lcc@{}}
\toprule
\textbf{Hyperparameter} & \textbf{Qwen2.5-7B} & \textbf{Gemma2-9B} \\
\midrule
Learning rate & $5\times10^{-6}$ & $8\times10^{-6}$ \\
Training epochs & 10 & 20 \\
Max sequence length & 2500 & 2500 \\
Per-device batch size & 4 & 4 \\
Gradient accumulation steps & 4 & 4 \\
LR scheduler & Cosine & Cosine \\
Warmup & 0.05 ratio & 20 steps \\
Optimizer & AdamW & AdamW \\
Precision & bfloat16 & bfloat16 \\
LoRA rank & 8 & 16 \\
LoRA alpha & 16 & 32 \\
LoRA dropout & 0.05 & 0.10 \\
LoRA target modules & all  & all \\
\bottomrule
\end{tabularx}
\caption{Core fine-tuning hyperparameters for the two open-source LLM extractors.}
\label{tab:finetune_hyperparams}
\end{table}
\subsection{Encoder and Prompting}

\paragraph{Encoder baselines.}
We compared against RoBERTa~\cite{liu2019roberta} and
SecureBERT~\cite{aghaei2022securebert} as segment-level encoder baselines.
We reformulate extraction as binary classification: given a candidate statement
and its section context, formatted as \textit{candidate statement}
\texttt{[SEP]} \textit{section context}, the encoder predicts whether the
candidate is a formalizable security goal. Candidate statements are generated
from annotated protocol chunks by splitting sentences, bullet items, table rows,
and local sentence combinations. Candidates aligned with gold goals are labeled
positive, and unmatched candidates are labeled negative. During inference, all
candidates are scored, thresholded, and deduplicated to obtain the final
extracted goal set. The two baselines use the same pipeline and differ only in
the pretrained checkpoint.

\begin{table}[ht]
\centering
\footnotesize
\setlength{\tabcolsep}{3pt}
\renewcommand{\arraystretch}{1.08}
\begin{tabularx}{\columnwidth}{@{}l>{\centering\arraybackslash}X>{\centering\arraybackslash}X@{}}
\toprule
\textbf{Hyperparameter} & \textbf{RoBERTa} & \textbf{SecureBERT} \\
\midrule
Base model & \texttt{roberta-base} & \makecell{\texttt{ehsanaghaei/}\\\texttt{SecureBERT}} \\
Max length & 512 & 512 \\
Batch size & 16 & 16 \\
Learning rate & $2\times10^{-5}$ & $1\times10^{-5}$ \\
Epochs & 10 & 12 \\
Weight decay & 0.01 & 0.01 \\
Warmup ratio & 0.1 & 0.1 \\
Focal $\gamma$ & 2.0 & 2.0 \\
Focal $\alpha$ & 0.25 & 0.25 \\
Inference threshold & 0.25 & 0.25 \\
Model-selection metric & $F_{0.5}$ & $F_{1.5}$ \\
\bottomrule
\end{tabularx}
\caption{Core hyperparameters for the encoder baselines.}
\label{tab:encoder_hyperparams}
\end{table}
We used focal loss for both encoders and fixed the random seed to 42. For each
baseline, we selected the checkpoint with the best validation score according
to the model-selection metric in Table~\ref{tab:encoder_hyperparams}.

\paragraph{Prompting setup.}
We evaluated prompting-based LLMs under zero-shot and few-shot settings. In
zero-shot prompting, the model received only the task instruction and the input
protocol chunk. In few-shot prompting, we prepended three manually constructed
demonstration examples to specify the output format and clarify the boundary
between formalizable security goals and non-goal protocol text. The few-shot
demonstrations include brief exclusion rationales, but the model was always
instructed to return JSON-only outputs. These examples were synthetic and were
not taken from any SecGoal benchmark document or fine-tuning data. All
prompting-based LLMs were evaluated on the full SecGoal benchmark.
\subsection{Retrieval and Formalization Setup}

We implemented the formalization stage using RAGFlow~\cite{ragflow2024}. The
chat model was GPT-5.4, and the embedding model was
\texttt{text-embedding-large}. The knowledge base contained the formal property
taxonomy, template schema, protocol flow definitions, and supporting protocol
context. Documents were processed using RAGFlow's default ``General'' parsing
mode.

We enabled Reasoning Flow and Keyword Analysis and kept the remaining RAGFlow
parameters at their default settings. The same retrieval and generation
configuration was used for all protocols in the formalization evaluation.

\begin{table}[ht]
    \centering
    \begingroup
    \footnotesize
    \setlength{\tabcolsep}{2pt}
    \renewcommand{\arraystretch}{0.95}
    \begin{tabularx}{\columnwidth}{@{}lccc>{\raggedright\arraybackslash}X@{}}
\toprule
\multicolumn{1}{c}{\textbf{Protocol}} &
\makecell{\textbf{Goal}\\\textbf{Stmts.}} &
\textbf{Props.} &
\makecell{\textbf{Goal}\\\textbf{Share}} &
\multicolumn{1}{c}{\textbf{Ref.}} \\
\midrule
        TLS 1.3      & 29  & 12  & 2.3\%  &~\cite{bhargavan2017verified,cremers2017comprehensive} \\
        IKEv2        & 20  & 8   & 1.3\%  &~\cite{gazdag2021formal,cremers2011key} \\
        Kerberos5    & 33  & 9   & 2.2\%  &~\cite{butler2006formal} \\
        OAuth2.0     & 51  & 7   & 3.6\%  &~\cite{fett2016comprehensive} \\
        5G-AKA       & 46  & 11  & 1.8\%  &~\cite{basin2018formal,cremers2019component} \\
        PQXDH        & 16  & 7   & 8.9\%  &~\cite{bhargavan2024formal} \\
        FIDO2        & 23  & 15  & 7.2\%  &~\cite{feng2021formal,guan2022formal} \\
        QUIC         & 25  & 7   & 0.9\%  &~\cite{zhang2021formal} \\
        SSH          & 10  & 8   & 2.7\%  &~\cite{ylonen2006rfc} \\
        OPC-UA       & 28  & 6   & 3.5\%  &~\cite{diemunsch2025comprehensive} \\
        FDO          & 23  & 6   & 10.7\% &~\cite{bussa2023formal} \\
        EDHOC        & 33  & 17  & 4.8\%  &~\cite{jacomme2023comprehensive} \\
        SPDM         & 20  & 8   & 1.1\%  &~\cite{cremers2023formal} \\
        NSSK         & 19  & 5   & 17.4\% &~\cite{clark1997survey} \\
        Kao-Chow v1  & 15  & 5   & 13.2\% &~\cite{clark1997survey} \\
        \midrule
        \textbf{Total} & \textbf{391} & \textbf{131} & \textbf{2.5\%} & -- \\
        \bottomrule
    \end{tabularx}
    \endgroup
    \caption{SecGoal dataset statistics across 15 protocols before document-level splitting. Goal Share denotes the approximate share of protocol document text annotated as security goal statements.}
    \label{tab:dataset_stats}
\end{table}
\section{Dataset Construction and Documentation}
\subsection{Coverage and Sources}
Table~\ref{tab:dataset_stats} provides the protocol-level documentation for
SecGoal, including annotated security goal statements, associated formal
properties, the approximate share of protocol document text annotated as
security goal statements, and source references. We use these statistics to track
annotation coverage.

\subsection{Train-Test Split}
We used a protocol-level train--test split to evaluate cross-protocol
generalization. All annotated statements from the same protocol were assigned
to the same split, avoiding leakage from protocol-specific terminology,
message names, and security assumptions across training and testing. The test
set consists of five held-out protocols: Kao-Chow v1, NSSK, SPDM, EDHOC, and
OPC-UA\@. The remaining ten protocols were used for training and validation.
\subsection{Three-Phase Annotation Pipeline}

Section~\ref{sec:hitl} describes the three-phase annotation pipeline; here, we report the additional per-phase statistics omitted there. We used frontier LLMs to generate broad preliminary candidate pools for expert annotation. After pooling and de-duplicating candidates within each document, the recall-oriented generation stage yielded two candidate pools containing 1,980 and 2,054 candidate security goal statements, respectively, for human review.

During independent review, each annotator confirmed or rejected every candidate, added goal statements missed by the generator, and assigned one or more property types. Annotators worked from the protocol specification and the references in Table~\ref{tab:dataset_stats}, with no access to each other's labels. Across the two independent reviews, annotators recovered 87 omitted goal statements that were absent from the LLM-generated candidate pools. Before arbitration, the two annotators selected 317 and 305 statements, respectively, with 263 shared statements and a union of 359 statements; see Table~\ref{tab:iaa}.

The senior professor adjudicated the 96 non-shared statements and all uncertain cases under the same specification-grounded inclusion rule, reconciled boundary and label differences, and recovered 29 additional omitted goal statements after the two annotators completed their reviews. The final dataset contains 391 goal statements and 131 formal security properties.

\subsection{Annotation Guidelines}

A security goal statement is an intended security property that the protocol guarantees, that is, the \emph{what}, and must be expressible as a trace constraint in a symbolic model. We exclude operational logic, namely the \emph{how}: parsing, encoding, and interoperability requirements.

Modal verbs such as ``must'' and ``shall'' are not sufficient evidence of security relevance. We also exclude non-formalizable requirements, such as UI mandates and physical constraints, as well as over-inferred goals. For example, the TLS 1.3 \texttt{legacy\_version} check serves middlebox interoperability rather than downgrade protection.

Each statement maps to one or more property types under a many-to-many relation.
\begin{table}[ht]
\centering
\footnotesize
\setlength{\tabcolsep}{1.5pt}
\resizebox{\columnwidth}{!}{
\begin{tabular}{@{}lrrrcccc@{}}
\toprule
\textbf{Protocol} & \textbf{A1} & \textbf{A2} & \textbf{Shr.}
& \textbf{Stmt. F1} & \textbf{Jacc.}
& \makecell{\textbf{Prop.}\\{\bfseries\boldmath $\mu$-F1}}
& \textbf{$Cohen's \kappa$} \\
\midrule
5G-AKA       & 31 & 33 & 24 & 0.750 & 0.600 & 0.873 & 0.859 \\
EDHOC        & 30 & 29 & 28 & 0.949 & 0.903 & 0.976 & 0.973 \\
FDO          & 19 & 15 & 13 & 0.765 & 0.619 & 1.000 & 1.000 \\
FIDO2        & 20 & 20 & 17 & 0.850 & 0.739 & 1.000 & 1.000 \\
IKEv2        & 16 & 18 & 14 & 0.824 & 0.700 & 0.902 & 0.877 \\
Kao-Chow v1  & 11 & 13 & 10 & 0.833 & 0.714 & 1.000 & 1.000 \\
Kerberos5    & 27 & 25 & 22 & 0.846 & 0.733 & 0.894 & 0.882 \\
NSSK         & 17 & 17 & 16 & 0.941 & 0.889 & 1.000 & 1.000 \\
OAuth2.0     & 40 & 37 & 32 & 0.831 & 0.711 & 0.965 & 0.959 \\
OPC-UA       & 27 & 21 & 21 & 0.875 & 0.778 & 0.951 & 0.938 \\
PQXDH        & 11 & 12 & 10 & 0.870 & 0.769 & 0.966 & 0.957 \\
QUIC         & 17 & 15 & 11 & 0.688 & 0.524 & 1.000 & 1.000 \\
SPDM         & 18 & 15 & 13 & 0.788 & 0.650 & 0.917 & 0.895 \\
SSH          &  8 & 10 &  8 & 0.889 & 0.800 & 0.903 & 0.872 \\
TLS 1.3      & 25 & 25 & 24 & 0.960 & 0.923 & 1.000 & 1.000 \\
\midrule
\textbf{Mean} & -- & -- & --
& \textbf{0.844} & \textbf{0.737}
& \textbf{0.956} & \textbf{0.947} \\
\textbf{Total} & \textbf{317} & \textbf{305} & \textbf{263}
& \multicolumn{4}{c}{Jaccard$=73.26\%$,\ F1$=84.57\%$} \\
\bottomrule
\end{tabular}
}
\caption{Inter-annotator agreement over the SecGoal annotation set, reported by protocol before final arbitration. Statement F1 and Jaccard measure statement-selection agreement; Prop.\ $\mu$-F1 and $\kappa$ measure property-label agreement.}
\label{tab:iaa}
\end{table}
\subsection{Annotator Qualifications and IAA}
\label{app:iaa}

Both annotators are PhD researchers with experience in cryptographic protocol analysis and symbolic verification tools. A senior professor performed final arbitration. Inter-annotator agreement was computed before arbitration on normalized selected statements, with duplicates removed within each protocol. Annotator~1 selected 317 statements and Annotator~2 selected 305 statements; 263 were shared, yielding 84.57\% Statement Selection F1 and 73.26\% Jaccard similarity. On shared statements, property-label agreement reached 0.956 Prop.\ $\mu$-F1 and 0.947 Cohen's $\kappa$. Detailed protocol-level results are reported in Table~\ref{tab:iaa}.

\section{Metric Definitions and a Worked Example}
\label{sec:appendix_metrics}

\paragraph{Why extraction uses cross-granularity metrics.}
Security-goal extraction in SecGoal is not ordinary sentence retrieval: it is
the upstream filtering step for formal verification. Two failure modes therefore
matter at the same time. First, false-positive extractions increase expert
review cost and may induce spurious formal properties. Second, false negatives
can make a downstream security property unrecoverable, even if many other
sentences are correctly extracted. Standard statement-level F1 does not capture
this behavior because the relation between natural-language statements and
formal properties is many-to-many: one statement may encode multiple formal
properties, while several different statements may support the same property.
We therefore score extraction at two granularities. $P_{\mathrm{ext}}$ measures
whether the extracted statements are clean security goal evidence, whereas
$R_{\mathrm{prop}}$ measures whether the extracted evidence still covers the
formal properties needed for downstream analysis. EC-F1 combines these two
requirements and is used as the primary extraction metric.

\paragraph{Extraction metrics.}
For a protocol, let $\mathcal{P}_{GT}$ be the set of ground-truth formal
properties and let $\mathcal{E}(p)$ denote the gold security goal statements
that provide evidence for property $p \in \mathcal{P}_{GT}$. Let
$\hat{\mathcal{G}}$ be the normalized and deduplicated set of extracted
statements. An extracted statement is counted as a true positive if it is a
valid, specification-grounded security goal statement and matches a gold
evidential statement; otherwise it is counted as a false positive. Thus,
\[
P_{\mathrm{ext}}=\frac{TP}{TP+FP}.
\]
We mark a ground-truth property as hit when its required evidential
statement(s) are present in the true-positive extraction set; for properties
with alternative evidential statements, recovering any sufficient evidence
counts as a hit. Let
$\mathcal{P}_{hit}$ be the set of hit properties. Property-level recall is
\[
R_{\mathrm{prop}}=\frac{|\mathcal{P}_{hit}|}{|\mathcal{P}_{GT}|}.
\]
Finally, extraction-coverage F1 is the harmonic mean of these two quantities:
\[
\mathrm{EC\text{-}F1}=
\frac{2\cdot P_{\mathrm{ext}}\cdot R_{\mathrm{prop}}}
{P_{\mathrm{ext}}+R_{\mathrm{prop}}}.
\]

\paragraph{Worked extraction example.}
Assume a protocol has six ground-truth formal properties,
$|\mathcal{P}_{GT}|=6$. A model extracts eight statements. Five are valid
security goal statements and three are operational requirements, such as message
encoding rules or abort conditions. Thus $TP=5$, $FP=3$, and
$P_{\mathrm{ext}}=5/(5+3)=0.625$. The five true-positive statements cover only
five distinct formal properties: one statement covers two authentication
properties, two statements redundantly support the same secrecy property, and
one remaining ground-truth property is missed. Hence $\#Hit=5$ and
$R_{\mathrm{prop}}=5/6=0.833$. The resulting score is
\[
\mathrm{EC\text{-}F1}
=\frac{2\cdot0.625\cdot0.833}{0.625+0.833}
=0.714.
\]
This example illustrates why both dimensions are necessary. A model can recover
most formal properties while still producing many false positives, or it can be
precise but miss a property required for verification. EC-F1 rewards extractors
that maintain downstream property coverage without flooding the formalization
stage with non-goal text.

\paragraph{Formalization metrics.}
For RQ3, we evaluate the generated formal-property set
$\hat{\mathcal{P}}$ against the gold set $\mathcal{P}_{GT}$ at two levels.

\textbf{Property-level matching.}
Each generated property can be matched to at most one gold property. A match
requires identical \texttt{type} and \texttt{subtype}, agreement on the
property-specific role fields, and sufficient overlap on the core data fields.
The default core-field threshold is $\tau=0.8$. Role fields are selected by
property family: authentication properties compare the two participant roles
and attacker type; secrecy and privacy properties compare the relevant role
sets and attacker type; and special properties compare the subtype-specific
actor or role fields. Core data fields are also family-specific: authentication
compares agreement values, secrecy compares secret or session data, privacy
compares protected identity data, and special properties compare the relevant
parameters, compromised keys, session keys, or evidence fields.

Let $TP$ be the number of generated properties that are successfully matched,
$FP$ the number of unmatched generated properties, and $FN$ the number of
unmatched gold properties. We compute
\[
\begin{aligned}
P_{\mathrm{prop}} &= \frac{TP}{TP+FP},\\
R_{\mathrm{prop}} &= \frac{TP}{TP+FN},\\
F1_{\mathrm{prop}} &=
\frac{2P_{\mathrm{prop}}R_{\mathrm{prop}}}
{P_{\mathrm{prop}}+R_{\mathrm{prop}}}.
\end{aligned}
\]
For NSSK and Kao-Chow v1, core-field overlap is evaluated by strict symbolic
equality: terms must match exactly, so \texttt{dec(X)} is not automatically
normalized to \texttt{X}, and both core-field precision and recall must reach
$\tau$. For EDHOC, whose properties have more complex internal structure, the
same one-to-one matching protocol uses a coverage-based core-data criterion:
if type, subtype, and roles match, a generated property is counted as covering a
gold property when it covers the gold core data. In all protocols, unmatched
generated and gold properties affect property-level precision and recall,
respectively.

\textbf{Slot-level scoring.}
Slot-level metrics evaluate whether the fields inside matched properties are
filled correctly. We flatten each property into a set of slot tokens. For
example, an Authentication property with \texttt{asserter=R},
\texttt{subject=I}, and agreement values
\texttt{TH\_2}, \texttt{TH\_3}, and \texttt{TH\_4} yields
\texttt{asserter=R}, \texttt{subject=I}, and one
\texttt{agreementValues=TH\_i} token for each
$i\in\{2,3,4\}$. For a matched pair, slot true positives are the shared
tokens, slot false positives are generated-only tokens, and slot false
negatives are gold-only tokens:
\[
\begin{aligned}
P_{\mathrm{slot}} &=
\frac{TP_{\mathrm{slot}}}{TP_{\mathrm{slot}}+FP_{\mathrm{slot}}},\\
R_{\mathrm{slot}} &=
\frac{TP_{\mathrm{slot}}}{TP_{\mathrm{slot}}+FN_{\mathrm{slot}}},\\
F1_{\mathrm{slot}} &=
\frac{2P_{\mathrm{slot}}R_{\mathrm{slot}}}
{P_{\mathrm{slot}}+R_{\mathrm{slot}}}.
\end{aligned}
\]
Slot scores are computed over matched property pairs for all protocols. Unmatched
generated and gold properties are captured by property-level precision and
recall, respectively, rather than being double-counted as slot-level errors.

\section{Formal Property Taxonomy and Templates}
\label{sec:appendix_templates}

The AIFG formalization stage uses a structured knowledge base to bridge
natural-language protocol specifications and symbolic security-property
descriptions. This appendix summarizes the supported property taxonomy and the
JSON schema used to instantiate formalized goals.

\subsection{Property Taxonomy}
The current schema covers four families of formalizable security properties:
\begin{itemize}[leftmargin=*,nosep]
    \item \textbf{Secrecy}: Covers standard confidentiality, strong secrecy (indistinguishability), forward and backward secrecy, and post-compromise security variants~\cite{blanchet2004automatic, boyd2003protocols, cohn2016post}.
    \item \textbf{Authentication}: Supports the hierarchy of authentication specifications defined by Lowe~\cite{lowe1997hierarchy}, including Aliveness, Weak Agreement, Non-injective Agreement, and Injective Agreement.
    \item \textbf{Privacy}: Addresses user-centric protections such as Identity Protection, Anonymity, and Unlinkability~\cite{pfitzmann2010terminology}.
    \item \textbf{Special Properties}: Includes other properties such as Downgrade Prevention~\cite{bhargavan2016downgrade}, Key Compromise Impersonation (KCI) protection, session independence, and Non-Repudiation~\cite{boyd2003protocols}.
\end{itemize}

\subsection{Formal Property Schema}
To avoid turning the appendix into a prompt dump, we summarize the executable
schema in Table~\ref{tab:formal_schema_summary} and show a compact JSON
skeleton below. The complete JSON schema, including all subtype-specific
\texttt{\_usage\_guide} fields, is included in the released artifact.

\begin{center}
\scriptsize
\setlength{\tabcolsep}{3pt}
\renewcommand{\arraystretch}{1.12}
\begin{tabularx}{\columnwidth}{@{}
    >{\raggedright\arraybackslash}p{0.23\columnwidth}
    >{\raggedright\arraybackslash}X
@{}}
\toprule
\textbf{Family} & \textbf{Schema Summary} \\
\midrule
\textbf{Secrecy} &
\textit{Subtypes}: \texttt{standard\_secrecy}, \texttt{strong\_secrecy}, \texttt{forward\_secrecy}, \texttt{backward\_secrecy}, \texttt{weak\_pcs}, \texttt{full\_pcs}.
\newline \textit{Core slots}: protected/compromised data, roles, recovery or healing mechanism.
\newline \textit{Attacker}: passive or active, subtype-dependent. \\
\textbf{Authentication} &
\textit{Subtypes}: \texttt{aliveness}, \texttt{weak\_agreement}, \texttt{non\_injective\_agreement}, \texttt{injective\_agreement}.
\newline \textit{Core slots}: \texttt{asserter}, \texttt{subject}, \texttt{agreementValues}.
\newline \textit{Attacker}: active. \\
\textbf{Privacy} &
\textit{Subtypes}: \texttt{low\_level\_privacy}, \texttt{anonymity}, \texttt{unlinkability}.
\newline \textit{Core slots}: identity variables, pseudonyms, session identifiers, compared identities.
\newline \textit{Attacker}: passive or active. \\
\textbf{Special} &
\textit{Subtypes}: \texttt{downgrade\_protection}, \texttt{kci\_protection}, \texttt{session\_independence}, \texttt{non\_repudiation}.
\newline \textit{Core slots}: negotiated parameters, compromised keys, session keys, signature evidence.
\newline \textit{Attacker}: active. \\
\bottomrule
\end{tabularx}
\refstepcounter{table}
\label{tab:formal_schema_summary}
\par\vspace{0.25em}
{\footnotesize\textbf{Table~\thetable:} Summary of the formal property schema used by AIFG.\par}
\end{center}

\begin{tcolorbox}[
    title={Compact Formal Property Schema Skeleton},
    breakable,
    colback=gray!5,
    colframe=gray!45,
    fonttitle=\bfseries\small
]
\begin{lstlisting}[
    basicstyle=\ttfamily\scriptsize,
    breaklines=true,
    columns=fullflexible,
    keepspaces=true
]
[
  {
    "property": {
      "type": "Secrecy",
      "subtype": "[registered_secrecy_subtype]",
      "secretData": "[Variable_Name_From_Flow]",
      "sharedBetween": ["[RoleA]", "[RoleB]"],
      "attacker_type": "[passive | active]"
    }
  },
  {
    "property": {
      "type": "Authentication",
      "subtype": "[registered_authentication_subtype]",
      "asserter": "[Role_Who_Verifies]",
      "subject": "[Role_Being_Authenticated]",
      "agreementValues": ["[Nonce_or_Key_From_Flow]"],
      "attacker_type": "active"
    }
  },
  {
    "property": {
      "type": "Privacy",
      "subtype": "[registered_privacy_subtype]",
      "identity": "[Identity_or_Pseudonym_From_Flow]",
      "sessionIdentifiers": ["[Session_Value_From_Flow]"],
      "attacker_type": "[passive | active]"
    }
  },
  {
    "property": {
      "type": "Special",
      "subtype": "[registered_special_subtype]",
      "target": "[Negotiated_or_Evidence_Value]",
      "dependsOn": ["[Protocol_State_or_Key]"],
      "attacker_type": "active"
    }
  },
  {
    "property": {
      "type": "Secrecy",
      "subtype": "forward_secrecy",
      "compromisedKey": "[LongTermKey_Variable]",
      "protectedSessionData": "[SessionKey_or_Message_Variable]",
      "recoveryMechanism": "[ephemeral_key_exchange]",
      "roles": ["[RoleA]", "[RoleB]"],
      "attacker_type": "active"
    }
  },
  {
    "property": {
      "type": "Privacy",
      "subtype": "unlinkability",
      "unlinkableData": "[Pseudonym_or_SessionID_Variable]",
      "attacker_type": "[passive | active]"
    }
  },
  {
    "property": {
      "type": "Special",
      "subtype": "downgrade_protection",
      "negotiatedParams": ["[Param]"],
      "strongestMode": "[Highest_Security_Mode]",
      "attacker_type": "active"
    }
  },
  {
    "property": {
      "type": "Special",
      "subtype": "kci_protection",
      "victimRole": "[Role_Being_Tricked]",
      "compromisedKey": "[LongTermKey_Variable]",
      "attacker_type": "active"
    }
  }
]
\end{lstlisting}
\end{tcolorbox}

\section{Prompt Templates}
\label{sec:appendix_prompts}

We report compact prompt skeletons for readability. The full executable prompts
used in experiments, including the complete few-shot demonstrations and all
field-filling rules, are included in the released artifact.

\subsection{Stage I: Security Goal Extraction}
The extraction prompt defines a security goal as a protection guarantee rather
than a mechanical protocol step. The zero-shot setting uses only the task
definition and input text. The few-shot setting inserts three synthetic
demonstrations with brief exclusion rationales before the final JSON-only
output instruction.

\begin{tcolorbox}[
    title={Extraction Prompt Skeleton},
    breakable,
    colback=gray!5,
    colframe=gray!45,
    fonttitle=\bfseries\small
]
\begin{lstlisting}[
    basicstyle=\ttfamily\scriptsize,
    breaklines=true,
    columns=fullflexible,
    keepspaces=true
]
You are a formal verification expert. Extract verbatim sentences that express SECURITY GOALS from protocol specification text.

A security goal states a protection guarantee: what asset is protected, which entity is authenticated, what attack is prevented, or what state must hold after a session.

Do not extract sentences that only describe message flow, encoding, field formats, abort conditions, algorithm selection, interoperability, UI/storage requirements, physical policies, or hardness assumptions.

If a sentence mixes a security guarantee with mechanism details, keep it only when the guarantee is explicit and independently meaningful.

Extract complete original sentences without rewriting, merging, or splitting them. Remove duplicate sentences while preserving document order.

Judge by semantic intent, not by MUST/SHALL keywords.

{few_shot_demonstrations_if_any}

Input Text:
{input_text}

Output only a JSON array of extracted sentences. If none qualify, output [].
\end{lstlisting}
\end{tcolorbox}

\subsection{Stage II: Security Goal Formalization}
The formalization prompt combines the extracted goal, retrieved protocol
context, protocol-flow definitions, and the formal schema from
Appendix~\ref{sec:appendix_templates}. It asks the model to discover the
security intent, ground abstract concepts to exact flow variables, and
instantiate the matching schema template.

\begin{tcolorbox}[
    title={Formalization Prompt Skeleton},
    breakable,
    colback=gray!5,
    colframe=gray!45,
    fonttitle=\bfseries\small
]
\begin{lstlisting}[
    basicstyle=\ttfamily\scriptsize,
    breaklines=true,
    columns=fullflexible,
    keepspaces=true
]
You are an expert Security Protocol Formalization Assistant.

Task:
Translate one natural-language security goal into strictly formatted JSON
security properties.

Inputs:
<formal_templates>{formal_templates}</formal_templates>
<output_schema>{output_schema}</output_schema>
<protocol_flow>{protocol_flow}</protocol_flow>
<symbol_inventory>{symbol_inventory}</symbol_inventory>
<role_alias_map>{role_alias_map}</role_alias_map>
<retrieved_context>{retrieved_context}</retrieved_context>
<security_goal>{security_goal}</security_goal>

Procedure:
1. Objective identification:
   Identify the concrete security objective(s) expressed or clearly entailed by
   {security_goal}, using {retrieved_context} only as supporting evidence.

2. Protocol grounding:
   Resolve roles through {role_alias_map}. Ground all protocol data using exact
   symbols from {symbol_inventory} or {protocol_flow}.

3. Template instantiation:
   Select the closest template from {formal_templates} for each distinct
   objective and fill only fields defined in {output_schema}.

Key constraints:
- Return only a valid JSON array. If no template applies, return [].
- Do not invent, rename, translate, or normalize protocol variables beyond
  {symbol_inventory}.
- Use canonical role names in all role fields; do not place role names inside
  agreementValues.
- If one goal implies multiple distinct objectives, output one JSON object per
  objective.
- For Authentication, asserter is the party receiving the assurance and subject
  is the authenticated peer. agreementValues contain only protocol data relevant
  to the agreement objective.
- For Secrecy, use the concrete protocol variable that represents the protected
  data or established secret. sharedBetween lists legitimate holders.
- For Privacy, use the privacy template when the protected data is an identity,
  credential identifier, endpoint identifier, metadata, or other private
  identifier.
- Use only fields defined in {output_schema}; omit helper or explanatory fields.

Return only the JSON array.
\end{lstlisting}
\end{tcolorbox}




\section{Detailed Results and Error Analysis}
The main evaluation reports aggregate answers to the three research questions.
This appendix uses the complete result tables to explain where those aggregate
trends come from. We focus on protocol-level heterogeneity, calibration effects
that are hidden by averages, and the concrete error mechanisms that connect
extraction quality to downstream property generation.

\subsection{RQ1 Complete Extraction Results}
\label{app:rq1_complete_extraction}

\paragraph{Protocol difficulty is highly uneven.}
Table~\ref{tab:rq1_complete_extraction} shows that the extraction problem is
not uniformly difficult across SecGoal. Prompted models perform best on compact
protocol descriptions whose security goals are close to the message-flow logic,
such as Kao-Chow v1, NSSK, PQXDH, SSH, and EDHOC. In these cases, the text
contains relatively little implementation or deployment material, so high
property recall is less often accompanied by large volumes of irrelevant
extractions. By contrast, QUIC, 5G-AKA, Kerberos5, FDO, OAuth2.0, OPC-UA, and
TLS~1.3 remain difficult even for stronger prompted models. Their
specifications contain extensive interoperability, state-machine, configuration,
and implementation requirements, which are security-adjacent but usually not
formalizable goals.

\paragraph{Few-shot prompting acts as local calibration, not task learning.}
The per-protocol rows also show that few-shot examples do not produce a stable
global decision boundary. They often improve precision on verbose protocols
such as TLS~1.3, but can reduce property recall on protocols where the gold
goals are sparse or expressed through protocol-specific mechanisms. For
example, several models become more selective after prompting while losing hits
on Kerberos5, PQXDH, or SPDM. This pattern suggests that in-context examples
mainly tune the model's extraction conservativeness; they do not provide the
protocol-specific formal-analysis knowledge needed to reliably distinguish
formalizable security goals from security-relevant mechanisms. Consistent with this
interpretation, a small number of few-shot outputs exhibit demonstration
leakage, where the model copies or paraphrases goals from the examples into the
target-protocol extraction; we treat these as false positives when they are not
supported by the target specification.

\subsection{RQ2 Complete Fine-tuning Results}
\label{app:rq2_complete_finetuning}

\paragraph{Fine-tuning changes the type of error.}
Table~\ref{tab:rq2_complete_finetuning} gives a more diagnostic view of the
fine-tuning gains reported in the main text. Encoder baselines are often
conservative: RoBERTa extracts few candidates and achieves perfect precision on
EDHOC, but misses many gold properties. SecureBERT improves coverage but
extracts more false positives. The fine-tuned LLMs occupy a different regime:
they retain broad property coverage while learning to suppress many
operational statements that prompted LLMs select. Thus the main effect of
SecGoal supervision is not simply to make the model extract more, but to change
which non-goal statements it rejects.

\paragraph{Fine-tuning does not guarantee complete goal coverage.}
The EDHOC rows expose a residual recall limitation that is hidden by the high
average $R_{\mathrm{prop}}$. Gemma2-9B-FT with section-level chunks and 1:3
downsampling reaches high precision, but still misses two gold properties.
Manual inspection shows that these misses trace to a subtle EDHOC goal
statement: transcript hashes (\texttt{TH\_2}, \texttt{TH\_3}, and
\texttt{TH\_4}) are used for key derivation and as additional authenticated
data. None of the evaluated prompted or fine-tuned extractors selected this
statement. This failure reflects two related challenges. First, even strong LLMs
still lack sufficient formal-analysis expertise to reliably recognize all
security goal statements described in protocol specifications. Second, this
statement is phrased through a concrete cryptographic mechanism rather than a
declarative security claim. Although this mechanism-level wording makes the
statement look like operational protocol logic, it is in fact a formalizable
security goal statement because it specifies the binding material that
underpins the intended authentication properties. Because all evaluated LLM
extractors miss this statement, the case shows that current LLMs remain
insufficient for complete security-goal extraction from protocol
specifications: fine-tuning improves selectivity, but it does not fully close
coverage gaps for mechanism-framed security goal statements.

\paragraph{The remaining errors are protocol-specific.}
The held-out protocols also reveal different residual precision failures. On
OPC-UA, the best fine-tuned model reaches full property recall but still
produces many false positives, reflecting the density of configuration,
negotiation, and deployment constraints in the specification. SPDM lies between
these cases: the model recovers all properties under the best setting, but
still includes redundant security-adjacent statements. These differences matter
because downstream formalization is affected differently by missed goals and by
redundant goals.

\subsection{RQ3 Complete Formalization Results}
\label{app:rq3_complete_formalization}

\paragraph{Input concision, not only correctness, affects property generation.}
Table~\ref{tab:rq3_complete_formalization} separates two factors that are
conflated in end-to-end use: whether the required goals are present, and
whether the corresponding goal statements are concise. Under
\textit{gold-minimal} inputs, AIFG receives concise goal statements aligned
with the target property set and therefore improves coverage, especially on
NSSK and SPDM, though precision remains sensitive to additional generated
properties and strict symbolic matching. Under \textit{expert-vetted} inputs,
false positives have already been removed, but the remaining extracted goal
statements are longer and less canonical. The resulting drop in precision shows
that correctness alone is insufficient: redundant or over-broad goal wording can
still lead AIFG to generate additional plausible structured properties.
These additional properties are not necessarily semantically wrong; in many
cases, they reflect small deviations from the minimal gold property set or from
the canonical slot choices used for evaluation.

\paragraph{Over-generation propagates differently at property and slot levels.}
The expert-vetted settings illustrate this distinction. AIFG generates more
properties than in the gold-minimal setting, lowering property precision across
all four protocols. However, the matched property pairs often retain stronger
slot-level scores: NSSK and Kao-Chow v1 reach perfect slot F1, while EDHOC and
SPDM retain 0.85 and 0.81 slot F1, respectively. This suggests that AIFG can
often ground roles, secrets, and agreement values accurately once a generated
property is matched to a gold property, even when it generates too many
structured properties overall. Manual inspection shows that non-perfect slot
F1 often reflects boundary errors within such matched properties. The compact
examples below illustrate both under-grounding and over-grounding after
property-level matching. These patterns motivate treating input concision as
an important objective, not only extraction correctness.

\begin{tcolorbox}[
    title={Slot-Level Error Examples in Matched Properties},
    breakable,
    colback=gray!5,
    colframe=gray!45,
    fonttitle=\bfseries\small
]
\begin{lstlisting}[
    basicstyle=\ttfamily\scriptsize,
    breaklines=true,
    columns=fullflexible,
    keepspaces=true
]
SPDM authentication (slot P/R/F1 = 1.000/0.875/0.933)
Gold:
{
  "type": "Authentication",
  "subtype": "non_injective_agreement",
  "asserter": "Requester",
  "subject": "Responder",
  "agreementValues": [
    "pkR",
    "digestR",
    "nonceChallengeI",
    "TVCA",
    "Tchallenge"
  ],
  "attacker_type": "active"
}

Generated:
{
  "type": "Authentication",
  "subtype": "non_injective_agreement",
  "asserter": "Requester",
  "subject": "Responder",
  "agreementValues": [
    "pkR",
    "nonceChallengeI",
    "TVCA",
    "Tchallenge"
  ],
  "attacker_type": "active"
}
Difference: missing "digestR" (under-grounding).

EDHOC explicit key confirmation (slot P/R/F1 = 0.800/1.000/0.889)
Gold:
{
  "type": "Authentication",
  "subtype": "non_injective_agreement",
  "asserter": "Responder",
  "subject": "Initiator",
  "agreementValues": [
    "method",
    "pkI",
    "pkR",
    "G_X",
    "G_Y",
    "PRK_4x3m"
  ]
}

Generated:
{
  "type": "Authentication",
  "subtype": "non_injective_agreement",
  "asserter": "Responder",
  "subject": "Initiator",
  "agreementValues": [
    "PRK_4x3m",
    "pkI",
    "pkR",
    "G_X",
    "G_Y",
    "method",
    "suitesI",
    "TH_4"
  ],
  "attacker_type": "active"
}
Difference: extra "suitesI" and "TH_4" (over-grounding).
\end{lstlisting}
\end{tcolorbox}

\clearpage
\onecolumn
\nolinenumbers

\begingroup
\setlength{\LTpre}{0pt}
\setlength{\LTpost}{6pt}
{\scriptsize
\setlength{\tabcolsep}{3pt}
\renewcommand{\arraystretch}{0.95}
\begin{longtable}{@{}llrcccccc@{}}
\caption{Complete RQ1 extraction results by model and protocol. ZS and FS denote zero-shot and few-shot prompting.}\label{tab:rq1_complete_extraction}\\
\toprule
\textbf{Model} & \textbf{Protocol} & \textbf{GT} & \textbf{ZS-$P_{\mathrm{ext}}$} & \textbf{ZS-$R_{\mathrm{prop}}$} & \textbf{ZS-EC-F1} & \textbf{FS-$P_{\mathrm{ext}}$} & \textbf{FS-$R_{\mathrm{prop}}$} & \textbf{FS-EC-F1} \\
\midrule
\endfirsthead
\caption[]{Complete RQ1 extraction results (continued).}\\
\toprule
\textbf{Model} & \textbf{Protocol} & \textbf{GT} & \textbf{ZS-$P_{\mathrm{ext}}$} & \textbf{ZS-$R_{\mathrm{prop}}$} & \textbf{ZS-EC-F1} & \textbf{FS-$P_{\mathrm{ext}}$} & \textbf{FS-$R_{\mathrm{prop}}$} & \textbf{FS-EC-F1} \\
\midrule
\endhead
\midrule
\multicolumn{9}{@{}l@{}}{\emph{Continued on next page}}\\
\endfoot
\bottomrule
\endlastfoot
Claude-Sonnet-4-6 & 5G-AKA & 11 & 0.099 & 1.000 & 0.180 & 0.105 & 1.000 & 0.191 \\
Claude-Sonnet-4-6 & EDHOC & 17 & 0.270 & 0.765 & 0.399 & 0.287 & 0.765 & 0.418 \\
Claude-Sonnet-4-6 & FDO & 6 & 0.125 & 1.000 & 0.222 & 0.138 & 1.000 & 0.243 \\
Claude-Sonnet-4-6 & FIDO2 & 15 & 0.182 & 1.000 & 0.309 & 0.194 & 1.000 & 0.324 \\
Claude-Sonnet-4-6 & IKEv2 & 8 & 0.131 & 1.000 & 0.232 & 0.131 & 1.000 & 0.232 \\
Claude-Sonnet-4-6 & Kao-Chow v1 & 5 & 0.394 & 1.000 & 0.565 & 0.429 & 1.000 & 0.600 \\
Claude-Sonnet-4-6 & Kerberos5 & 9 & 0.116 & 1.000 & 0.208 & 0.123 & 1.000 & 0.219 \\
Claude-Sonnet-4-6 & NSSK & 5 & 0.343 & 1.000 & 0.511 & 0.361 & 1.000 & 0.531 \\
Claude-Sonnet-4-6 & OAuth2.0 & 7 & 0.133 & 1.000 & 0.235 & 0.141 & 1.000 & 0.248 \\
Claude-Sonnet-4-6 & OPC-UA & 6 & 0.101 & 1.000 & 0.183 & 0.129 & 1.000 & 0.229 \\
Claude-Sonnet-4-6 & PQXDH & 7 & 0.293 & 1.000 & 0.453 & 0.333 & 1.000 & 0.500 \\
Claude-Sonnet-4-6 & QUIC & 7 & 0.076 & 1.000 & 0.140 & 0.076 & 1.000 & 0.141 \\
Claude-Sonnet-4-6 & SPDM & 8 & 0.214 & 1.000 & 0.353 & 0.185 & 1.000 & 0.312 \\
Claude-Sonnet-4-6 & SSH & 8 & 0.414 & 1.000 & 0.585 & 0.400 & 1.000 & 0.571 \\
Claude-Sonnet-4-6 & TLS 1.3 & 12 & 0.147 & 1.000 & 0.257 & 0.163 & 1.000 & 0.280 \\
\textbf{Claude-Sonnet-4-6} & \textbf{Average} & \textbf{8.7} & \textbf{0.203} & \textbf{0.984} & \textbf{0.322} & \textbf{0.213} & \textbf{0.984} & \textbf{0.336} \\
\midrule
DeepSeek-V4-Flash & 5G-AKA & 11 & 0.114 & 1.000 & 0.204 & 0.138 & 1.000 & 0.243 \\
DeepSeek-V4-Flash & EDHOC & 17 & 0.309 & 0.765 & 0.440 & 0.382 & 0.647 & 0.481 \\
DeepSeek-V4-Flash & FDO & 6 & 0.118 & 1.000 & 0.211 & 0.134 & 1.000 & 0.237 \\
DeepSeek-V4-Flash & FIDO2 & 15 & 0.353 & 1.000 & 0.522 & 0.492 & 1.000 & 0.659 \\
DeepSeek-V4-Flash & IKEv2 & 8 & 0.179 & 1.000 & 0.304 & 0.173 & 1.000 & 0.295 \\
DeepSeek-V4-Flash & Kao-Chow v1 & 5 & 0.471 & 1.000 & 0.640 & 0.421 & 1.000 & 0.593 \\
DeepSeek-V4-Flash & Kerberos5 & 9 & 0.131 & 1.000 & 0.231 & 0.156 & 1.000 & 0.270 \\
DeepSeek-V4-Flash & NSSK & 5 & 0.385 & 1.000 & 0.556 & 0.385 & 1.000 & 0.556 \\
DeepSeek-V4-Flash & OAuth2.0 & 7 & 0.145 & 1.000 & 0.254 & 0.161 & 1.000 & 0.277 \\
DeepSeek-V4-Flash & OPC-UA & 6 & 0.131 & 1.000 & 0.232 & 0.137 & 1.000 & 0.241 \\
DeepSeek-V4-Flash & PQXDH & 7 & 0.474 & 0.857 & 0.610 & 0.643 & 0.857 & 0.735 \\
DeepSeek-V4-Flash & QUIC & 7 & 0.080 & 1.000 & 0.148 & 0.110 & 1.000 & 0.199 \\
DeepSeek-V4-Flash & SPDM & 8 & 0.165 & 0.875 & 0.277 & 0.218 & 0.875 & 0.349 \\
DeepSeek-V4-Flash & SSH & 8 & 0.280 & 0.875 & 0.424 & 0.364 & 0.875 & 0.514 \\
DeepSeek-V4-Flash & TLS 1.3 & 12 & 0.142 & 0.833 & 0.243 & 0.217 & 1.000 & 0.357 \\
\textbf{DeepSeek-V4-Flash} & \textbf{Average} & \textbf{8.7} & \textbf{0.232} & \textbf{0.947} & \textbf{0.353} & \textbf{0.275} & \textbf{0.950} & \textbf{0.400} \\
\midrule
DeepSeek-V4-Pro & 5G-AKA & 11 & 0.124 & 1.000 & 0.220 & 0.141 & 1.000 & 0.247 \\
DeepSeek-V4-Pro & EDHOC & 17 & 0.338 & 0.765 & 0.468 & 0.448 & 0.765 & 0.565 \\
DeepSeek-V4-Pro & FDO & 6 & 0.137 & 1.000 & 0.242 & 0.143 & 1.000 & 0.250 \\
DeepSeek-V4-Pro & FIDO2 & 15 & 0.545 & 1.000 & 0.706 & 0.480 & 1.000 & 0.649 \\
DeepSeek-V4-Pro & IKEv2 & 8 & 0.126 & 1.000 & 0.224 & 0.156 & 1.000 & 0.269 \\
DeepSeek-V4-Pro & Kao-Chow v1 & 5 & 0.458 & 1.000 & 0.629 & 0.550 & 1.000 & 0.710 \\
DeepSeek-V4-Pro & Kerberos5 & 9 & 0.133 & 1.000 & 0.235 & 0.138 & 0.889 & 0.239 \\
DeepSeek-V4-Pro & NSSK & 5 & 0.385 & 1.000 & 0.556 & 0.429 & 1.000 & 0.600 \\
DeepSeek-V4-Pro & OAuth2.0 & 7 & 0.165 & 1.000 & 0.283 & 0.162 & 0.857 & 0.272 \\
DeepSeek-V4-Pro & OPC-UA & 6 & 0.130 & 1.000 & 0.230 & 0.157 & 1.000 & 0.271 \\
DeepSeek-V4-Pro & PQXDH & 7 & 0.429 & 0.857 & 0.571 & 0.643 & 0.857 & 0.735 \\
DeepSeek-V4-Pro & QUIC & 7 & 0.070 & 1.000 & 0.131 & 0.087 & 1.000 & 0.161 \\
DeepSeek-V4-Pro & SPDM & 8 & 0.133 & 0.875 & 0.230 & 0.188 & 1.000 & 0.316 \\
DeepSeek-V4-Pro & SSH & 8 & 0.296 & 0.875 & 0.443 & 0.320 & 0.875 & 0.469 \\
DeepSeek-V4-Pro & TLS 1.3 & 12 & 0.146 & 1.000 & 0.255 & 0.186 & 1.000 & 0.313 \\
\textbf{DeepSeek-V4-Pro} & \textbf{Average} & \textbf{8.7} & \textbf{0.241} & \textbf{0.958} & \textbf{0.361} & \textbf{0.282} & \textbf{0.950} & \textbf{0.404} \\
\midrule
Gemini-3-Flash-Preview & 5G-AKA & 11 & 0.125 & 1.000 & 0.222 & 0.183 & 1.000 & 0.309 \\
Gemini-3-Flash-Preview & EDHOC & 17 & 0.349 & 0.765 & 0.480 & 0.441 & 0.765 & 0.559 \\
Gemini-3-Flash-Preview & FDO & 6 & 0.142 & 1.000 & 0.248 & 0.161 & 1.000 & 0.277 \\
Gemini-3-Flash-Preview & FIDO2 & 15 & 0.283 & 1.000 & 0.441 & 0.381 & 1.000 & 0.552 \\
Gemini-3-Flash-Preview & IKEv2 & 8 & 0.148 & 1.000 & 0.257 & 0.189 & 1.000 & 0.318 \\
Gemini-3-Flash-Preview & Kao-Chow v1 & 5 & 0.480 & 1.000 & 0.649 & 0.400 & 1.000 & 0.571 \\
Gemini-3-Flash-Preview & Kerberos5 & 9 & 0.133 & 1.000 & 0.235 & 0.162 & 0.778 & 0.269 \\
Gemini-3-Flash-Preview & NSSK & 5 & 0.414 & 1.000 & 0.585 & 0.500 & 1.000 & 0.667 \\
Gemini-3-Flash-Preview & OAuth2.0 & 7 & 0.176 & 1.000 & 0.299 & 0.168 & 1.000 & 0.288 \\
Gemini-3-Flash-Preview & OPC-UA & 6 & 0.121 & 1.000 & 0.216 & 0.198 & 1.000 & 0.331 \\
Gemini-3-Flash-Preview & PQXDH & 7 & 0.476 & 0.857 & 0.612 & 0.538 & 0.857 & 0.661 \\
Gemini-3-Flash-Preview & QUIC & 7 & 0.096 & 1.000 & 0.175 & 0.099 & 1.000 & 0.180 \\
Gemini-3-Flash-Preview & SPDM & 8 & 0.193 & 0.875 & 0.316 & 0.260 & 0.875 & 0.401 \\
Gemini-3-Flash-Preview & SSH & 8 & 0.435 & 0.875 & 0.581 & 0.625 & 0.875 & 0.729 \\
Gemini-3-Flash-Preview & TLS 1.3 & 12 & 0.216 & 1.000 & 0.356 & 0.229 & 1.000 & 0.372 \\
\textbf{Gemini-3-Flash-Preview} & \textbf{Average} & \textbf{8.7} & \textbf{0.252} & \textbf{0.958} & \textbf{0.378} & \textbf{0.302} & \textbf{0.943} & \textbf{0.432} \\
\midrule
Gemini-3-Pro-Preview & 5G-AKA & 11 & 0.160 & 0.909 & 0.272 & 0.209 & 1.000 & 0.345 \\
Gemini-3-Pro-Preview & EDHOC & 17 & 0.464 & 0.765 & 0.578 & 0.581 & 0.765 & 0.661 \\
Gemini-3-Pro-Preview & FDO & 6 & 0.189 & 1.000 & 0.319 & 0.210 & 1.000 & 0.347 \\
Gemini-3-Pro-Preview & FIDO2 & 15 & 0.439 & 1.000 & 0.610 & 0.521 & 1.000 & 0.685 \\
Gemini-3-Pro-Preview & IKEv2 & 8 & 0.280 & 1.000 & 0.438 & 0.282 & 1.000 & 0.440 \\
Gemini-3-Pro-Preview & Kao-Chow v1 & 5 & 0.500 & 1.000 & 0.667 & 0.526 & 1.000 & 0.690 \\
Gemini-3-Pro-Preview & Kerberos5 & 9 & 0.179 & 0.889 & 0.299 & 0.241 & 0.778 & 0.368 \\
Gemini-3-Pro-Preview & NSSK & 5 & 0.435 & 1.000 & 0.606 & 0.435 & 1.000 & 0.606 \\
Gemini-3-Pro-Preview & OAuth2.0 & 7 & 0.183 & 1.000 & 0.310 & 0.214 & 1.000 & 0.353 \\
Gemini-3-Pro-Preview & OPC-UA & 6 & 0.180 & 1.000 & 0.306 & 0.222 & 1.000 & 0.364 \\
Gemini-3-Pro-Preview & PQXDH & 7 & 0.500 & 0.857 & 0.632 & 0.636 & 0.857 & 0.730 \\
Gemini-3-Pro-Preview & QUIC & 7 & 0.112 & 1.000 & 0.201 & 0.148 & 1.000 & 0.257 \\
Gemini-3-Pro-Preview & SPDM & 8 & 0.317 & 0.875 & 0.465 & 0.286 & 0.875 & 0.431 \\
Gemini-3-Pro-Preview & SSH & 8 & 0.529 & 0.875 & 0.660 & 0.563 & 0.875 & 0.685 \\
Gemini-3-Pro-Preview & TLS 1.3 & 12 & 0.267 & 1.000 & 0.421 & 0.321 & 1.000 & 0.486 \\
\textbf{Gemini-3-Pro-Preview} & \textbf{Average} & \textbf{8.7} & \textbf{0.316} & \textbf{0.945} & \textbf{0.452} & \textbf{0.360} & \textbf{0.943} & \textbf{0.496} \\
\midrule
GLM-5 & 5G-AKA & 11 & 0.169 & 1.000 & 0.289 & 0.225 & 1.000 & 0.367 \\
GLM-5 & EDHOC & 17 & 0.468 & 0.765 & 0.580 & 0.628 & 0.765 & 0.690 \\
GLM-5 & FDO & 6 & 0.167 & 1.000 & 0.286 & 0.243 & 1.000 & 0.391 \\
GLM-5 & FIDO2 & 15 & 0.440 & 1.000 & 0.611 & 0.485 & 1.000 & 0.653 \\
GLM-5 & IKEv2 & 8 & 0.218 & 1.000 & 0.358 & 0.360 & 1.000 & 0.529 \\
GLM-5 & Kao-Chow v1 & 5 & 0.455 & 1.000 & 0.625 & 0.600 & 1.000 & 0.750 \\
GLM-5 & Kerberos5 & 9 & 0.177 & 0.889 & 0.295 & 0.278 & 0.556 & 0.370 \\
GLM-5 & NSSK & 5 & 0.400 & 1.000 & 0.571 & 0.385 & 1.000 & 0.556 \\
GLM-5 & OAuth2.0 & 7 & 0.214 & 1.000 & 0.352 & 0.211 & 0.857 & 0.339 \\
GLM-5 & OPC-UA & 6 & 0.156 & 1.000 & 0.270 & 0.247 & 1.000 & 0.396 \\
GLM-5 & PQXDH & 7 & 0.524 & 0.857 & 0.650 & 0.300 & 0.571 & 0.393 \\
GLM-5 & QUIC & 7 & 0.139 & 1.000 & 0.244 & 0.161 & 1.000 & 0.278 \\
GLM-5 & SPDM & 8 & 0.292 & 1.000 & 0.452 & 0.316 & 0.875 & 0.464 \\
GLM-5 & SSH & 8 & 0.563 & 0.875 & 0.685 & 0.636 & 0.875 & 0.737 \\
GLM-5 & TLS 1.3 & 12 & 0.246 & 1.000 & 0.395 & 0.356 & 1.000 & 0.525 \\
\textbf{GLM-5} & \textbf{Average} & \textbf{8.7} & \textbf{0.308} & \textbf{0.959} & \textbf{0.444} & \textbf{0.362} & \textbf{0.900} & \textbf{0.496} \\
\midrule
GPT-5.4 & 5G-AKA & 11 & 0.112 & 1.000 & 0.202 & 0.119 & 1.000 & 0.212 \\
GPT-5.4 & EDHOC & 17 & 0.270 & 0.765 & 0.399 & 0.370 & 0.765 & 0.499 \\
GPT-5.4 & FDO & 6 & 0.142 & 1.000 & 0.248 & 0.168 & 1.000 & 0.288 \\
GPT-5.4 & FIDO2 & 15 & 0.277 & 1.000 & 0.434 & 0.304 & 1.000 & 0.467 \\
GPT-5.4 & IKEv2 & 8 & 0.133 & 1.000 & 0.235 & 0.151 & 1.000 & 0.262 \\
GPT-5.4 & Kao-Chow v1 & 5 & 0.407 & 1.000 & 0.579 & 0.435 & 1.000 & 0.606 \\
GPT-5.4 & Kerberos5 & 9 & 0.123 & 1.000 & 0.220 & 0.124 & 1.000 & 0.220 \\
GPT-5.4 & NSSK & 5 & 0.361 & 1.000 & 0.531 & 0.353 & 1.000 & 0.522 \\
GPT-5.4 & OAuth2.0 & 7 & 0.152 & 1.000 & 0.264 & 0.145 & 1.000 & 0.254 \\
GPT-5.4 & OPC-UA & 6 & 0.118 & 1.000 & 0.211 & 0.128 & 1.000 & 0.227 \\
GPT-5.4 & PQXDH & 7 & 0.323 & 1.000 & 0.488 & 0.333 & 1.000 & 0.500 \\
GPT-5.4 & QUIC & 7 & 0.072 & 1.000 & 0.135 & 0.077 & 1.000 & 0.143 \\
GPT-5.4 & SPDM & 8 & 0.136 & 0.875 & 0.236 & 0.261 & 0.875 & 0.402 \\
GPT-5.4 & SSH & 8 & 0.316 & 1.000 & 0.480 & 0.429 & 1.000 & 0.600 \\
GPT-5.4 & TLS 1.3 & 12 & 0.160 & 1.000 & 0.276 & 0.186 & 1.000 & 0.313 \\
\textbf{GPT-5.4} & \textbf{Average} & \textbf{8.7} & \textbf{0.207} & \textbf{0.976} & \textbf{0.329} & \textbf{0.239} & \textbf{0.976} & \textbf{0.368} \\
\midrule
Qwen3.5-Plus & 5G-AKA & 11 & 0.137 & 1.000 & 0.241 & 0.184 & 0.909 & 0.306 \\
Qwen3.5-Plus & EDHOC & 17 & 0.475 & 0.765 & 0.586 & 0.596 & 0.765 & 0.670 \\
Qwen3.5-Plus & FDO & 6 & 0.139 & 1.000 & 0.245 & 0.189 & 1.000 & 0.318 \\
Qwen3.5-Plus & FIDO2 & 15 & 0.366 & 1.000 & 0.536 & 0.512 & 1.000 & 0.677 \\
Qwen3.5-Plus & IKEv2 & 8 & 0.158 & 0.875 & 0.268 & 0.250 & 0.875 & 0.389 \\
Qwen3.5-Plus & Kao-Chow v1 & 5 & 0.480 & 1.000 & 0.649 & 0.500 & 1.000 & 0.667 \\
Qwen3.5-Plus & Kerberos5 & 9 & 0.175 & 0.889 & 0.292 & 0.227 & 0.667 & 0.338 \\
Qwen3.5-Plus & NSSK & 5 & 0.435 & 1.000 & 0.606 & 0.417 & 1.000 & 0.588 \\
Qwen3.5-Plus & OAuth2.0 & 7 & 0.183 & 1.000 & 0.310 & 0.179 & 1.000 & 0.303 \\
Qwen3.5-Plus & OPC-UA & 6 & 0.176 & 1.000 & 0.300 & 0.237 & 1.000 & 0.383 \\
Qwen3.5-Plus & PQXDH & 7 & 0.588 & 0.857 & 0.698 & 0.875 & 0.857 & 0.866 \\
Qwen3.5-Plus & QUIC & 7 & 0.124 & 1.000 & 0.220 & 0.127 & 1.000 & 0.225 \\
Qwen3.5-Plus & SPDM & 8 & 0.250 & 0.875 & 0.389 & 0.345 & 0.875 & 0.495 \\
Qwen3.5-Plus & SSH & 8 & 0.474 & 0.875 & 0.615 & 0.533 & 0.875 & 0.663 \\
Qwen3.5-Plus & TLS 1.3 & 12 & 0.197 & 1.000 & 0.330 & 0.297 & 1.000 & 0.458 \\
\textbf{Qwen3.5-Plus} & \textbf{Average} & \textbf{8.7} & \textbf{0.291} & \textbf{0.942} & \textbf{0.419} & \textbf{0.364} & \textbf{0.922} & \textbf{0.490} \\
\midrule
Gemma2-9B-Instruct & 5G-AKA & 11 & 0.063 & 0.455 & 0.110 & 0.064 & 0.455 & 0.112 \\
Gemma2-9B-Instruct & EDHOC & 17 & 0.259 & 0.647 & 0.370 & 0.291 & 0.647 & 0.402 \\
Gemma2-9B-Instruct & FDO & 6 & 0.053 & 0.833 & 0.100 & 0.068 & 0.833 & 0.126 \\
Gemma2-9B-Instruct & FIDO2 & 15 & 0.078 & 0.800 & 0.142 & 0.096 & 0.867 & 0.173 \\
Gemma2-9B-Instruct & IKEv2 & 8 & 0.057 & 1.000 & 0.109 & 0.060 & 0.875 & 0.112 \\
Gemma2-9B-Instruct & Kao-Chow v1 & 5 & 0.350 & 1.000 & 0.519 & 0.320 & 1.000 & 0.485 \\
Gemma2-9B-Instruct & Kerberos5 & 9 & 0.082 & 0.667 & 0.146 & 0.087 & 0.667 & 0.153 \\
Gemma2-9B-Instruct & NSSK & 5 & 0.345 & 1.000 & 0.513 & 0.406 & 1.000 & 0.578 \\
Gemma2-9B-Instruct & OAuth2.0 & 7 & 0.087 & 1.000 & 0.161 & 0.091 & 1.000 & 0.167 \\
Gemma2-9B-Instruct & OPC-UA & 6 & 0.082 & 1.000 & 0.151 & 0.077 & 1.000 & 0.143 \\
Gemma2-9B-Instruct & PQXDH & 7 & 0.200 & 0.857 & 0.324 & 0.214 & 0.571 & 0.312 \\
Gemma2-9B-Instruct & QUIC & 7 & 0.030 & 0.571 & 0.058 & 0.036 & 0.857 & 0.069 \\
Gemma2-9B-Instruct & SPDM & 8 & 0.078 & 0.750 & 0.141 & 0.107 & 0.750 & 0.188 \\
Gemma2-9B-Instruct & SSH & 8 & 0.100 & 0.625 & 0.172 & 0.129 & 0.500 & 0.205 \\
Gemma2-9B-Instruct & TLS 1.3 & 12 & 0.047 & 0.500 & 0.086 & 0.089 & 0.750 & 0.159 \\
\textbf{Gemma2-9B-Instruct} & \textbf{Average} & \textbf{8.7} & \textbf{0.127} & \textbf{0.780} & \textbf{0.207} & \textbf{0.142} & \textbf{0.785} & \textbf{0.226} \\
\midrule
Qwen2.5-7B-Instruct & 5G-AKA & 11 & 0.162 & 0.818 & 0.270 & 0.216 & 0.364 & 0.271 \\
Qwen2.5-7B-Instruct & EDHOC & 17 & 0.477 & 0.765 & 0.588 & 0.424 & 0.706 & 0.530 \\
Qwen2.5-7B-Instruct & FDO & 6 & 0.194 & 1.000 & 0.325 & 0.200 & 0.667 & 0.308 \\
Qwen2.5-7B-Instruct & FIDO2 & 15 & 0.149 & 0.800 & 0.252 & 0.189 & 0.800 & 0.305 \\
Qwen2.5-7B-Instruct & IKEv2 & 8 & 0.048 & 0.625 & 0.090 & 0.051 & 0.250 & 0.085 \\
Qwen2.5-7B-Instruct & Kao-Chow v1 & 5 & 0.500 & 1.000 & 0.667 & 0.333 & 1.000 & 0.500 \\
Qwen2.5-7B-Instruct & Kerberos5 & 9 & 0.141 & 0.556 & 0.225 & 0.167 & 0.333 & 0.222 \\
Qwen2.5-7B-Instruct & NSSK & 5 & 0.348 & 1.000 & 0.516 & 0.545 & 1.000 & 0.706 \\
Qwen2.5-7B-Instruct & OAuth2.0 & 7 & 0.141 & 1.000 & 0.247 & 0.198 & 1.000 & 0.330 \\
Qwen2.5-7B-Instruct & OPC-UA & 6 & 0.185 & 1.000 & 0.313 & 0.182 & 1.000 & 0.308 \\
Qwen2.5-7B-Instruct & PQXDH & 7 & 0.300 & 0.714 & 0.423 & 0.200 & 0.571 & 0.296 \\
Qwen2.5-7B-Instruct & QUIC & 7 & 0.030 & 0.857 & 0.057 & 0.131 & 0.714 & 0.222 \\
Qwen2.5-7B-Instruct & SPDM & 8 & 0.154 & 0.750 & 0.255 & 0.143 & 0.500 & 0.222 \\
Qwen2.5-7B-Instruct & SSH & 8 & 0.250 & 0.500 & 0.333 & 0.300 & 0.375 & 0.333 \\
Qwen2.5-7B-Instruct & TLS 1.3 & 12 & 0.085 & 0.833 & 0.154 & 0.153 & 0.667 & 0.248 \\
\textbf{Qwen2.5-7B-Instruct} & \textbf{Average} & \textbf{8.7} & \textbf{0.211} & \textbf{0.815} & \textbf{0.314} & \textbf{0.229} & \textbf{0.663} & \textbf{0.326} \\
\end{longtable}
}

{\scriptsize
\setlength{\tabcolsep}{3pt}
\renewcommand{\arraystretch}{1.05}
\begin{longtable}{@{}llrrrrrrrr@{}}
\caption{Complete RQ2 test-set extraction results. In Gemma2-9B-FT variants, \textit{sent.} denotes sentence-window chunking and \textit{sect.} denotes section-level chunking.}\label{tab:rq2_complete_finetuning}\\
\toprule
\textbf{Model} & \textbf{Protocol} & \textbf{GT} & \textbf{Ext.} & \textbf{TP} & \textbf{FP} & \textbf{\#Hit} & \textbf{$P_{\mathrm{ext}}$} & \textbf{$R_{\mathrm{prop}}$} & \textbf{EC-F1} \\
\midrule
\endfirsthead
\caption[]{Complete RQ2 test-set extraction results (continued).}\\
\toprule
\textbf{Model} & \textbf{Protocol} & \textbf{GT} & \textbf{Ext.} & \textbf{TP} & \textbf{FP} & \textbf{\#Hit} & \textbf{$P_{\mathrm{ext}}$} & \textbf{$R_{\mathrm{prop}}$} & \textbf{EC-F1} \\
\midrule
\endhead
\midrule
\multicolumn{10}{@{}l@{}}{\emph{Continued on next page}}\\
\endfoot
\bottomrule
\endlastfoot
RoBERTa & EDHOC & 17 & 7 & 7 & 0 & 8 & 1.000 & 0.471 & 0.640 \\
RoBERTa & Kao-Chow v1 & 5 & 10 & 6 & 4 & 4 & 0.600 & 0.800 & 0.686 \\
RoBERTa & NSSK & 5 & 11 & 6 & 5 & 4 & 0.545 & 0.800 & 0.649 \\
RoBERTa & OPC-UA & 6 & 4 & 1 & 3 & 1 & 0.250 & 0.167 & 0.200 \\
RoBERTa & SPDM & 8 & 6 & 3 & 3 & 7 & 0.500 & 0.875 & 0.636 \\
\textbf{RoBERTa} & \textbf{Average} & \textbf{8.2} & \textbf{7.6} & \textbf{4.6} & \textbf{3.0} & \textbf{4.8} & \textbf{0.579} & \textbf{0.622} & \textbf{0.562} \\
\midrule
SecureBERT & EDHOC & 17 & 16 & 11 & 5 & 9 & 0.688 & 0.529 & 0.598 \\
SecureBERT & Kao-Chow v1 & 5 & 21 & 13 & 8 & 5 & 0.619 & 1.000 & 0.765 \\
SecureBERT & NSSK & 5 & 22 & 14 & 8 & 5 & 0.636 & 1.000 & 0.778 \\
SecureBERT & OPC-UA & 6 & 18 & 7 & 11 & 5 & 0.389 & 0.833 & 0.530 \\
SecureBERT & SPDM & 8 & 12 & 8 & 4 & 5 & 0.667 & 0.625 & 0.645 \\
\textbf{SecureBERT} & \textbf{Average} & \textbf{8.2} & \textbf{17.8} & \textbf{10.6} & \textbf{7.2} & \textbf{5.8} & \textbf{0.600} & \textbf{0.798} & \textbf{0.663} \\
\midrule
Qwen2.5-7B-FT & EDHOC & 17 & 32 & 19 & 13 & 13 & 0.594 & 0.765 & 0.668 \\
Qwen2.5-7B-FT & Kao-Chow v1 & 5 & 13 & 10 & 3 & 5 & 0.769 & 1.000 & 0.870 \\
Qwen2.5-7B-FT & NSSK & 5 & 11 & 8 & 3 & 5 & 0.727 & 1.000 & 0.842 \\
Qwen2.5-7B-FT & OPC-UA & 6 & 52 & 15 & 37 & 6 & 0.288 & 1.000 & 0.448 \\
Qwen2.5-7B-FT & SPDM & 8 & 11 & 6 & 5 & 5 & 0.545 & 0.625 & 0.583 \\
\textbf{Qwen2.5-7B-FT} & \textbf{Average} & \textbf{8.2} & \textbf{23.8} & \textbf{11.6} & \textbf{12.2} & \textbf{6.8} & \textbf{0.585} & \textbf{0.878} & \textbf{0.682} \\
\midrule
Gemma2-9B-FT (sent., none) & EDHOC & 17 & 25 & 19 & 6 & 10 & 0.760 & 0.588 & 0.663 \\
Gemma2-9B-FT (sent., none) & Kao-Chow v1 & 5 & 11 & 8 & 3 & 5 & 0.727 & 1.000 & 0.842 \\
Gemma2-9B-FT (sent., none) & NSSK & 5 & 15 & 8 & 7 & 5 & 0.533 & 1.000 & 0.696 \\
Gemma2-9B-FT (sent., none) & OPC-UA & 6 & 30 & 15 & 15 & 5 & 0.500 & 0.833 & 0.625 \\
Gemma2-9B-FT (sent., none) & SPDM & 8 & 11 & 7 & 4 & 6 & 0.636 & 0.750 & 0.689 \\
\textbf{Gemma2-9B-FT (sent., none)} & \textbf{Average} & \textbf{8.2} & \textbf{18.4} & \textbf{11.4} & \textbf{7.0} & \textbf{6.2} & \textbf{0.631} & \textbf{0.834} & \textbf{0.703} \\
\midrule
Gemma2-9B-FT (sent., 1:3) & EDHOC & 17 & 31 & 23 & 8 & 11 & 0.742 & 0.647 & 0.691 \\
Gemma2-9B-FT (sent., 1:3) & Kao-Chow v1 & 5 & 13 & 8 & 5 & 5 & 0.615 & 1.000 & 0.762 \\
Gemma2-9B-FT (sent., 1:3) & NSSK & 5 & 14 & 9 & 5 & 5 & 0.643 & 1.000 & 0.783 \\
Gemma2-9B-FT (sent., 1:3) & OPC-UA & 6 & 38 & 19 & 19 & 6 & 0.500 & 1.000 & 0.667 \\
Gemma2-9B-FT (sent., 1:3) & SPDM & 8 & 14 & 11 & 3 & 7 & 0.786 & 0.875 & 0.828 \\
\textbf{Gemma2-9B-FT (sent., 1:3)} & \textbf{Average} & \textbf{8.2} & \textbf{22.0} & \textbf{14.0} & \textbf{8.0} & \textbf{6.8} & \textbf{0.657} & \textbf{0.904} & \textbf{0.746} \\
\midrule
Gemma2-9B-FT (sect., none) & EDHOC & 17 & 30 & 22 & 8 & 14 & 0.733 & 0.824 & 0.776 \\
Gemma2-9B-FT (sect., none) & Kao-Chow v1 & 5 & 10 & 7 & 3 & 5 & 0.700 & 1.000 & 0.824 \\
Gemma2-9B-FT (sect., none) & NSSK & 5 & 9 & 5 & 4 & 3 & 0.556 & 0.600 & 0.577 \\
Gemma2-9B-FT (sect., none) & OPC-UA & 6 & 32 & 15 & 17 & 6 & 0.469 & 1.000 & 0.638 \\
Gemma2-9B-FT (sect., none) & SPDM & 8 & 10 & 6 & 4 & 7 & 0.600 & 0.875 & 0.712 \\
\textbf{Gemma2-9B-FT (sect., none)} & \textbf{Average} & \textbf{8.2} & \textbf{18.2} & \textbf{11.0} & \textbf{7.2} & \textbf{7.0} & \textbf{0.612} & \textbf{0.860} & \textbf{0.705} \\
\midrule
Gemma2-9B-FT (sect., 1:3) & EDHOC & 17 & 33 & 27 & 6 & 15 & 0.818 & 0.882 & 0.849 \\
Gemma2-9B-FT (sect., 1:3) & Kao-Chow v1 & 5 & 9 & 8 & 1 & 5 & 0.889 & 1.000 & 0.941 \\
Gemma2-9B-FT (sect., 1:3) & NSSK & 5 & 10 & 6 & 4 & 5 & 0.600 & 1.000 & 0.750 \\
Gemma2-9B-FT (sect., 1:3) & OPC-UA & 6 & 36 & 16 & 20 & 6 & 0.444 & 1.000 & 0.615 \\
Gemma2-9B-FT (sect., 1:3) & SPDM & 8 & 19 & 11 & 8 & 8 & 0.579 & 1.000 & 0.733 \\
\textbf{Gemma2-9B-FT (sect., 1:3)} & \textbf{Average} & \textbf{8.2} & \textbf{21.4} & \textbf{13.6} & \textbf{7.8} & \textbf{7.8} & \textbf{0.666} & \textbf{0.976} & \textbf{0.778} \\
\end{longtable}
}

\begin{table}[!htbp]
\centering
\scriptsize
\setlength{\tabcolsep}{3.2pt}
\renewcommand{\arraystretch}{1.08}
\resizebox{\textwidth}{!}{
\begin{tabular}{@{}llccccccccccc@{}}
\toprule
\textbf{Protocol} & \textbf{Input} & \textbf{Gold} & \textbf{Gen.} & \textbf{TP} & \textbf{FP} & \textbf{FN} & \multicolumn{3}{c}{\textbf{Property}} & \multicolumn{3}{c}{\textbf{Slot}} \\
\cmidrule(lr){8-10}\cmidrule(lr){11-13}
 & & & & & & & \textbf{P} & \textbf{R} & \textbf{F1} & \textbf{P} & \textbf{R} & \textbf{F1} \\
\midrule
NSSK & gold-minimal & 5.00$\pm$0.00 & 6.33$\pm$0.94 & 5.00$\pm$0.00 & 2.00$\pm$0.00 & 0.00$\pm$0.00 & 0.71$\pm$0.00 & 1.00$\pm$0.00 & 0.83$\pm$0.00 & 0.93$\pm$0.09 & 1.00$\pm$0.00 & 0.96$\pm$0.05 \\
NSSK & expert-vetted & 5.00$\pm$0.00 & 8.67$\pm$0.94 & 5.00$\pm$0.00 & 3.67$\pm$0.94 & 0.00$\pm$0.00 & 0.58$\pm$0.06 & 1.00$\pm$0.00 & 0.74$\pm$0.05 & 1.00$\pm$0.00 & 1.00$\pm$0.00 & 1.00$\pm$0.00 \\
Kao-Chow v1 & gold-minimal & 5.00$\pm$0.00 & 5.00$\pm$0.00 & 3.67$\pm$0.94 & 1.33$\pm$0.94 & 1.33$\pm$0.94 & 0.73$\pm$0.19 & 0.73$\pm$0.19 & 0.73$\pm$0.19 & 0.89$\pm$0.08 & 1.00$\pm$0.00 & 0.94$\pm$0.04 \\
Kao-Chow v1 & expert-vetted & 5.00$\pm$0.00 & 13.33$\pm$0.94 & 5.00$\pm$0.00 & 8.33$\pm$0.94 & 0.00$\pm$0.00 & 0.38$\pm$0.03 & 1.00$\pm$0.00 & 0.55$\pm$0.03 & 1.00$\pm$0.00 & 1.00$\pm$0.00 & 1.00$\pm$0.00 \\
EDHOC & gold-minimal & 17.00$\pm$0.00 & 20.00$\pm$0.00 & 14.67$\pm$0.47 & 5.33$\pm$0.47 & 2.33$\pm$0.47 & 0.73$\pm$0.02 & 0.86$\pm$0.03 & 0.79$\pm$0.03 & 0.85$\pm$0.00 & 0.89$\pm$0.00 & 0.87$\pm$0.00 \\
EDHOC & expert-vetted & 17.00$\pm$0.00 & 34.33$\pm$2.36 & 13.00$\pm$0.00 & 21.33$\pm$2.36 & 4.00$\pm$0.00 & 0.38$\pm$0.03 & 0.76$\pm$0.00 & 0.51$\pm$0.02 & 0.83$\pm$0.01 & 0.89$\pm$0.00 & 0.85$\pm$0.01 \\
SPDM & gold-minimal & 8.00$\pm$0.00 & 12.33$\pm$1.70 & 7.33$\pm$0.94 & 6.00$\pm$1.41 & 0.67$\pm$0.94 & 0.55$\pm$0.08 & 0.92$\pm$0.12 & 0.69$\pm$0.09 & 0.77$\pm$0.08 & 0.98$\pm$0.02 & 0.84$\pm$0.06 \\
SPDM & expert-vetted & 8.00$\pm$0.00 & 21.00$\pm$0.82 & 5.67$\pm$0.47 & 15.67$\pm$0.94 & 2.33$\pm$0.47 & 0.27$\pm$0.03 & 0.71$\pm$0.06 & 0.39$\pm$0.04 & 0.73$\pm$0.06 & 0.95$\pm$0.01 & 0.81$\pm$0.05 \\
\bottomrule
\end{tabular}
}
\caption{Complete RQ3 formalization results under the property-level and slot-level evaluation, reported as mean $\pm$ standard deviation over three runs.}
\label{tab:rq3_complete_formalization}
\end{table}

\endgroup
\end{document}